\documentclass{aa}  

\usepackage{graphicx}
\usepackage{txfonts}
\usepackage{natbib}
\bibpunct{(}{)}{;}{a}{}{,} % to follow the A&A style

\usepackage{pifont}% http://ctan.org/pkg/pifont
\newcommand{\cmark}{\ding{51}}%
\newcommand{\xmark}{\ding{55}}%

\usepackage{color}
\newcommand{\arcs}{^{\prime\prime}}
\usepackage{hyperref}
\hypersetup{breaklinks=True,
    colorlinks=true,
    linkcolor=blue,
    citecolor=blue,
    filecolor=cyan,      
    urlcolor=blue,
    pdftitle={The ``Maggie'' filament},
    pdfauthor={J. Syed}}

\begin{document}

   \title{The ``Maggie'' filament:\\Physical properties of a giant atomic cloud}

   \subtitle{}

   \author{J. Syed\inst{\ref{inst_mpia}}
          \and J. D. Soler\inst{\ref{inst_mpia}}
          \and H. Beuther\inst{\ref{inst_mpia}}
          \and Y. Wang\inst{\ref{inst_mpia}}
          \and S. Suri\inst{\ref{inst_vien},\ref{inst_mpia}}
          \and J. D. Henshaw\inst{\ref{inst_mpia}}
          \and M. Riener\inst{\ref{inst_mpia}}
          \and S. Bialy\inst{\ref{inst_cfa}}
          \and S. Rezaei Kh.\inst{\ref{inst_mpia},\ref{inst_chalm}}
          \and J. M. Stil\inst{\ref{inst_calg}}
          \and P. F. Goldsmith\inst{\ref{inst_jpl}}
          \and M. R. Rugel\inst{\ref{inst_mpifr}}
          \and S. C. O. Glover\inst{\ref{inst_ita}}
          \and R. S. Klessen\inst{\ref{inst_ita},\ref{inst_unihd}}
          \and J. Kerp\inst{\ref{inst_arge}}
          \and J. S. Urquhart\inst{\ref{inst_kent}}
          \and J. Ott\inst{\ref{inst_nrao}}
          \and N. Roy\inst{\ref{inst_ind}}
          \and N. Schneider\inst{\ref{inst_koeln}}
          \and R. J. Smith\inst{\ref{inst_man}}
          \and S. N. Longmore\inst{\ref{inst_liv}}
          \and H. Linz\inst{\ref{inst_mpia}}
          }

   \institute{Max-Planck-Institut f\"ur Astronomie,
              K\"onigstuhl 17, 69117 Heidelberg, Germany\\
              \email{syed@mpia.de}\label{inst_mpia}
              \and University of Vienna, Department of Astrophysics, T\"urkenschanzstrasse 17, 1180 Vienna, Austria\label{inst_vien}
              \and Harvard Smithsonian Center for Astrophysics, 60 Garden Street, Cambridge, MA, 02138, USA\label{inst_cfa}
              \and Chalmers University of Technology, Department of Space, Earth and Environment, 412 93 Gothenburg, Sweden\label{inst_chalm}
              \and Department of Physics and Astronomy, The University of Calgary, 2500 University Drive NW, Calgary AB T2N 1N4, Canada\label{inst_calg}
              \and Jet Propulsion Laboratory, California Institute of Technology, 4800 Oak Grove Drive, Pasadena, CA 91109, USA\label{inst_jpl}
              \and Max-Planck-Institut f\"ur Radioastronomie, Auf dem H\"ugel 69, 53121 Bonn, Germany\label{inst_mpifr}
              \and Universit\"at Heidelberg, Zentrum für Astronomie, Institut für Theoretische Astrophysik, Albert-Ueberle-Str. 2, 69120 Heidelberg, Germany\label{inst_ita}
              \and Universit\"at Heidelberg, Interdisziplin\"ares Zentrum für Wissenschaftliches Rechnen, INF 205, 69120 Heidelberg, Germany\label{inst_unihd}
              \and Argelander-Institut f\"ur Astronomie, Auf dem H\"ugel 71, 53121 Bonn, Germany\label{inst_arge}
              \and Centre for Astrophysics and Planetary Science, University of Kent, Canterbury CT2 7NH, UK\label{inst_kent}
              \and National Radio Astronomy Observatory, PO Box O, 1003 Lopezville Road, Socorro, NM 87801, USA\label{inst_nrao}
              \and Department of Physics, Indian Institute of Science, Bengaluru 560012, India\label{inst_ind}
              \and I. Physik. Institut, University of Cologne, Z\"ulpicher Str. 77, 50937 Cologne, Germany\label{inst_koeln}
              \and Jodrell Bank Centre for Astrophysics, Department of Physics and Astronomy, University of Manchester, Oxford Road, Manchester M13 9PL, UK\label{inst_man}
              \and Astrophysics Research Institute, Liverpool John Moores University, IC2, Liverpool Science Park, 146 Brownlow Hill, Liverpool L3 5RF, UK\label{inst_liv}
             }

   \date{Received , xxxx; accepted , xxxx}

% \abstract{}{}{}{}{} 
% 5 {} token are mandatory
 
  \abstract
   {The atomic phase of the interstellar medium plays a key role in the formation process of molecular clouds. Due to the line-of-sight confusion in the Galactic plane that is associated with its ubiquity, atomic hydrogen emission has been challenging to study.}
   {We investigate the physical properties of the ``Maggie'' filament, a large-scale filament identified in \ion{H}{i} emission at line-of-sight velocities, $\varv_{\mathrm{LSR}}\sim -54\rm\,km\,s^{-1}$.}
   {Employing the high-angular resolution data from the THOR survey, we are able to study \ion{H}{i} emission features at negative $\varv_{\mathrm{LSR}}$ velocities without any line-of-sight confusion due to the kinematic distance ambiguity in the first Galactic quadrant. In order to investigate the kinematic structure, we decompose the emission spectra using the automated Gaussian fitting algorithm \textsc{GaussPy+}.}
   {We identify one of the largest, coherent, mostly atomic HI filaments in the Milky Way. The giant atomic filament Maggie, with a total length of $1.2\pm 0.1\rm\,kpc$, is not detected in most other tracers, and does not show signs of active star formation. At a kinematic distance of 17\,kpc, Maggie is situated below (by $\approx$500\,pc) but parallel to the Galactic HI disk and is trailing the predicted location of the Outer Arm by $5-10\rm\,km\,s^{-1}$ in longitude-velocity space. The centroid velocity exhibits a smooth gradient of less than $\pm 3\rm\,km\,s^{-1}\,(10\,pc)^{-1}$ and a coherent structure to within $\pm 6\rm\,km\,s^{-1}$. The line widths of $\sim$10$\rm\,km\,s^{-1}$ along the spine of the filament are dominated by non-thermal effects. After correcting for optical depth effects, the mass of Maggie's dense spine is estimated to be $7.2\substack{+2.5 \\ -1.9}\times 10^5\,M_{\odot}$. The mean number density of the filament is $\sim$4$\rm\,cm^{-3}$, which is best explained by the filament being a mix of cold and warm neutral gas. In contrast to molecular filaments, the turbulent Mach number and velocity structure function suggest that Maggie is driven by transonic to moderately supersonic velocities that are likely associated with the Galactic potential rather than being subject to the effects of self-gravity or stellar feedback. The column density PDF displays a log-normal shape around a mean of $\langle N_{\ion{H}{i}}\rangle = 4.8\times 10^{20}\rm\,cm^{-2}$, thus reflecting the absence of dominating effects of gravitational contraction.}
   {While Maggie's origin remains unclear, we hypothesize that Maggie could be the first in a class of atomic clouds that are the precursors of giant molecular filaments.}


   \titlerunning{The ``Maggie'' filament}
   \authorrunning{J. Syed et al.}
   \maketitle
%
%-------------------------------------------------------------------
\section{Introduction}

   Stars form in the cold, dense interiors of molecular clouds. The physical properties of these clouds therefore set the initial conditions under which star formation takes place. A key question in understanding the star formation process as part of the global interstellar matter cycle addresses the formation of large-scale molecular clouds out of the diffuse atomic phase of the interstellar medium \citep[ISM; for a review see][]{2001RvMP...73.1031F,Draine2011,2016SAAS...43...85K}. However, the atomic ISM and its dynamical properties are still observationally poorly constrained. 
   
   The ISM has a hierarchical structure and facilitates the formation of filaments that are governed by the Galactic potential on a large scale. \citet{2020A&A...642A.163S} present a network of \ion{H}{i} filaments already evident in the diffuse atomic phase of the ISM that is structured mostly parallel to the Galactic plane. Only in a few cases is the orientation of the filaments locally no longer dictated by the overall drag of the Galactic disk but rather is determined by the effects of stellar feedback and strong magnetic fields.
   \citet{2020A&A...642A.163S} identify a unique \ion{H}{i} filament that they have named ``Maggie'', after the largest river in Colombia, the R\'io Magdalena. Maggie is shown to be a highly elongated filamentary cloud (see Fig.~\ref{fig:HI_overview}) extending over $\sim$4$^{\circ}$ on the sky in Galactic longitude. Given its central velocity of $\varv_{\mathrm{LSR}}\approx -54\rm\,km\,s^{-1}$ and assuming circular motion, Maggie is located approximately $17\rm\,kpc$ away from us and has a length of more than $1\rm\,kpc$.
   
   In this paper, we present a detailed study of the Maggie filament and aim to understand its physical nature. Is Maggie the precursor of large-scale molecular filaments? Are the physical and kinematic properties of filamentary molecular clouds inherited from an atomic counterpart in the diffuse ISM? Giant molecular filaments are to date the largest coherent entities identified in the Milky Way and have been subject of many recent studies investigating the evolution of large filaments with respect to the global dynamics of the Milky Way on the one hand and local stellar feedback on the other hand \citep{2010ApJ...719L.185J,2014A&A...568A..73R,2014ApJ...797...53G,2015MNRAS.450.4043W,2016ApJS..226....9W,2016A&A...590A.131A,2015ApJ...815...23Z,2018ApJ...864..153Z,2020A&A...641A..53W}.
   
   The highly filamentary infrared dark cloud (IRDC) ``Nessie'', first identified in \citet{2010ApJ...719L.185J}, is argued to be one of the first in a class of filaments whose morphology is likely governed by the structural dynamics of the Galaxy. \citet{2014ApJ...797...53G} dubbed this type of filaments ``bones'' of the Milky Way -- highly filamentary molecular clouds whose formation, evolution, and shape could be closely linked to the global spiral structure of the Galaxy.
   
   \citet{2018ApJ...864..153Z} utilized a standardized approach to re-analyze a set of Galactic filaments identified in the literature. In doing that, it is possible to draw meaningful conclusions and reliably compare physical properties between filaments rather than being subject to the systematics of selection criteria and different methodology. They found in their sample of filaments that the giant molecular filaments (GMFs), first identified in \citet{2014A&A...568A..73R} and \citet{2016A&A...590A.131A}, have the highest masses ($\sim$10$^5\,M_{\odot}$) while exhibiting the lowest column densities and star-forming activity, making them good candidates to be the immediate descendants of atomic ISM structures.
   
   The GMFs are first identified as near and mid-infrared extinction features with a spatial extent of $\approx$100$\rm\,pc$. \citet{2014A&A...568A..73R} and \citet{2016A&A...590A.131A} then confirm velocity contiguity of the filaments via \element[][13]{CO} emission. Giant molecular filaments are not only associated with spiral arms but are also located in inter-arm regions.
   
   The recently discovered Radcliffe wave \citep{2020Natur.578..237A} is a coherent 2.7\,kpc long association of local molecular cloud complexes. It appears to be undulating above and below the Galactic midplane, and its three-dimensional shape (in position-position-position space) is well described by a damped sinusoidal wave. The Radcliffe wave provides a framework for future studies of molecular cloud formation and evolution with respect to the Galactic dynamics of the Milky Way.
   
   Simulations find that giant molecular clouds often form as large filaments, and their formation can be related to the dynamics of the galaxy and position with respect to the spiral arm potentials.
   \citet{2014MNRAS.441.1628S} use simulations of a four-armed spiral galaxy to investigate the formation of molecular gas structures. They find that high-density filaments tend to form in spiral arms while lower-density gas resides in long inter-arm spurs that are stretched by galactic shear. While the simulations account for the chemical evolution of the gas, they do not include the effects of self-gravity of the gas or stellar feedback.
   
   Including both stellar feedback and self-gravity, smoothed particle hydrodynamics simulations of a high-resolution section of a spiral galaxy indicate that galactic shear between spiral arms plays a critical role in the formation of highly filamentary structures \citep{2015MNRAS.447.3390D,2016MNRAS.458.3667D}. \citet{2017MNRAS.470.4261D} expand this analysis to the time evolution of these filaments, finding that giant molecular clouds tend to sustain their large filamentary shape before entering the spiral arms, where they are prone to being disrupted by local events of star formation and stellar feedback.
   
   The CloudFactory simulations presented in \citet{2020MNRAS.492.1594S} strongly suggest that spiral arms and differential rotation tend to arrange molecular clouds in long filaments that are likely to fragment while regions of clustered feedback randomizes the orientation of them. Molecular filaments that are dominated by galactic-scale forces also exhibit low internal velocity gradients and are tightly confined to the galactic plane. Filaments formed in regions of higher turbulence due to supernova (SN) feedback show a range of orientations with respect to the galactic plane and are more widely distributed.
   
   The systematic search for atomic counterparts to molecular clouds poses a challenging task. By means of the 21cm-line of \ion{H}{i} emission it is generally possible to probe large atomic clouds in the ISM but it has proven difficult to identify them as clearly defined objects in the inner Galactic plane. Traditionally, \ion{H}{i} clouds are then either observed at high Galactic latitudes \citep[e.g.][]{2016ApJ...821..117K} or they must have velocities significantly different than those imposed by the Galactic rotation \citep[e.g.][]{1997ARA&A..35..217W,2018MNRAS.474..289W}.
   
   As the analysis of \ion{H}{i} emission in the Galactic plane within the solar circle suffers from the kinematic distance ambiguity, any \ion{H}{i} cloud with $\varv_{\mathrm{LSR}}\gtrsim0\rm\,km\,s^{-1}$ (in the first Galactic quadrant) might be the product of blending foreground and background components that correspond to the same line-of-sight velocity. However, at negative $\varv_{\mathrm{LSR}}$ velocities any \ion{H}{i} structure can be identified with less line-of-sight confusion\footnote{vice versa in the fourth Galactic quadrant.} \citep[see e.g.][]{1993A&A...275...67B}.
   
    Additionally, the physical properties of atomic hydrogen are not straightforward to derive from emission studies alone.
    In thermal pressure equilibrium, theoretical considerations based on ISM heating and cooling processes predict two stable phases of atomic hydrogen at the observed pressures in the ISM, namely the cold neutral medium (CNM) and warm neutral medium \citep[WNM;][]{1969ApJ...155L.149F,1977ApJ...218..148M,2003ApJ...587..278W,2019ApJ...881..160B}. Observations of \ion{H}{i} emission are thus generally attributed to both CNM and WNM, which have significantly different physical properties. The CNM is observed to have temperatures $\lesssim 300\rm\,K$ and number densities of $\gtrsim n_{\mathrm{min,CNM}}=10\rm\,cm^{-3}$ while the thermally stable WNM exceeds temperatures of $\sim$5000$\rm\,K$ with number densities $\lesssim n_{\mathrm{max,WNM}}=0.1\rm\,cm^{-3}$ \citep{2003ApJ...586.1067H,2009ARA&A..47...27K}. For densities between $n_{\mathrm{min,CNM}}$ and $n_{\mathrm{max,WNM}}$, the gas is thermally unstable (denoted by UNM: unstable neutral medium) and it will move toward a stable CNM or WNM branch under isobaric density perturbations \citep{1965ApJ...142..531F}.
    
    Throughout the scope of this paper, we use the analytic model presented in \citet{2003ApJ...587..278W} as a standard reference. According to this, the two stable atomic hydrogen phases can coexist in thermal equilibrium over a narrow range of pressures, $P_{\mathrm{min}}\leq P\leq P_{\mathrm{max}}$, that is governed by the total heating and cooling rates of the interstellar gas (see e.g. the phase diagrams shown in Fig.~7 in \citealt{2003ApJ...587..278W}). The dominant heating process under typical ISM conditions is photoelectric (PE) heating from dust grains and polycyclic aromatic hydrocarbons (PAHs). The WNM is mainly cooled by Ly$\alpha$ emission while the most efficient cooling mechanism of the CNM is metal-line fine-structure emission primarily excited by collisions with electrons and neutral hydrogen atoms. Assuming a constant ratio of the cosmic-ray (or X-ray) ionization rate, $\zeta$, to the intensity of the UV interstellar radiation field, $I_{\mathrm{UV}}$, PE heating is proportional to $I_{\mathrm{UV}}$. If $I_{\mathrm{UV}}$ is then increased (decreased), the density at which metal-line cooling balances heating must increase (decrease) in the same manner. As a result, the pressure range at which both WNM and CNM occur shifts to higher or lower values in proportion to $I_{\mathrm{UV}}$.
    
    Furthermore, the metallicity $Z^{\prime}$ and dust abundance are crucial parameters for the thermal balance. If the metallicity is moderately lowered, the equilibrium pressure range dictating the WNM and CNM properties decreases since the dust-to-gas ratio and associated PE heating rate decreases more rapidly with decreasing $Z^{\prime}$ than does the metal-line cooling\footnote{Here we do not consider the effect of cosmic ray heating that becomes the dominant heating mechanism at metallicities $Z^{\prime}\lesssim0.1$ \citep[see Fig.~2 in][]{2019ApJ...881..160B}.} \citep[][see their Fig.~4]{2019ApJ...881..160B}.
    
    In an attempt to observationally isolate the CNM from the bistable emission, \ion{H}{i} self-absorption \citep[HISA;][]{2000ApJ...540..851G,2003ApJ...585..823L,2020A&A...634A.139W,2020A&A...642A..68S} is a viable method to study \ion{H}{i} clouds in the inner Milky Way but it heavily depends on the presence of sufficient background emission.
    
    The Maggie filament was discovered in \citet{2020A&A...642A.163S} via \ion{H}{i} emission in The \ion{H}{i}/OH Recombination line survey of the inner Milky Way \citep[THOR;][]{2016A&A...595A..32B,2020A&A...634A..83W}. It is identified as a large coherent filament using a Hessian matrix approach. This method allows to systematically identify filamentary structures by their spatial curvature that emerge as second derivative signatures in the \ion{H}{i} emission maps of the THOR survey.
    
    We follow up on this discovery and derive the physical properties of the Maggie filament in the subsequent analysis. This paper is organized as follows. We introduce the observations and data in Sect.~\ref{sec:obsandmethods} and outline the method we use for the spectral decomposition of the \ion{H}{i} emission. We present the fundamental properties of Maggie and show the results of the kinematics and column density in Sect.~\ref{sec:results}. In Sect.~\ref{sec:discussion} we examine Maggie for molecular counterparts and explore the possible formation process of Maggie. We also discuss the implications of the velocity and column density structure. Finally, we review the relationship between Maggie and giant molecular filaments and draw our conclusions in Sect.~\ref{sec:conclusions}.

%--------------------------------------------------------------------
\section{Observation and methods}\label{sec:obsandmethods}

\subsection{HI 21 cm line and continuum}\label{sec:observation}
   
   In the following analysis we employ the \ion{H}{i} and 1.4\,GHz continuum data from the THOR survey. A complete overview of the THOR survey and the data products is given in \citet{2016A&A...595A..32B} and \citet{2020A&A...634A..83W}. We investigate the kinematics based on the \ion{H}{i} emission data, where the 1.4 GHz continuum emission has been subtracted from the THOR \ion{H}{i} observation during data reduction. The final \ion{H}{i} emission data (THOR-\ion{H}{i}) have been obtained from observations with the Very Large Array (VLA) in C- and D-configuration, as well as single dish observations from the Greenbank Telescope. The data have an angular resolution of $40\arcs$ and the noise in emission-free channels is $\sim$4$\rm\,K$ at the spectral resolution of $1.5\rm\,km\,s^{-1}$. The covered velocities range from $-113$ to $+163\rm\,km\,s^{-1}$.
   
   In Sect.~\ref{sec:coldens_mass} we correct for the optical depth that we measured against discrete continuum sources. We therefore select the THOR \ion{H}{i}+continuum data that consist of VLA C-array configuration data only (THOR-only). THOR-only data have a higher angular resolution of $\sim$14\arcsec, making them suitable to study absorption against discrete continuum sources. Since this data set consists of interferometric observations only, large-scale \ion{H}{i} emission is effectively filtered out. This large-scale filtering due to the interferometer can be exploited when only discrete continuum sources are of interest.
   
   \begin{figure*}
      \centering
        \includegraphics[width=1.0\textwidth]{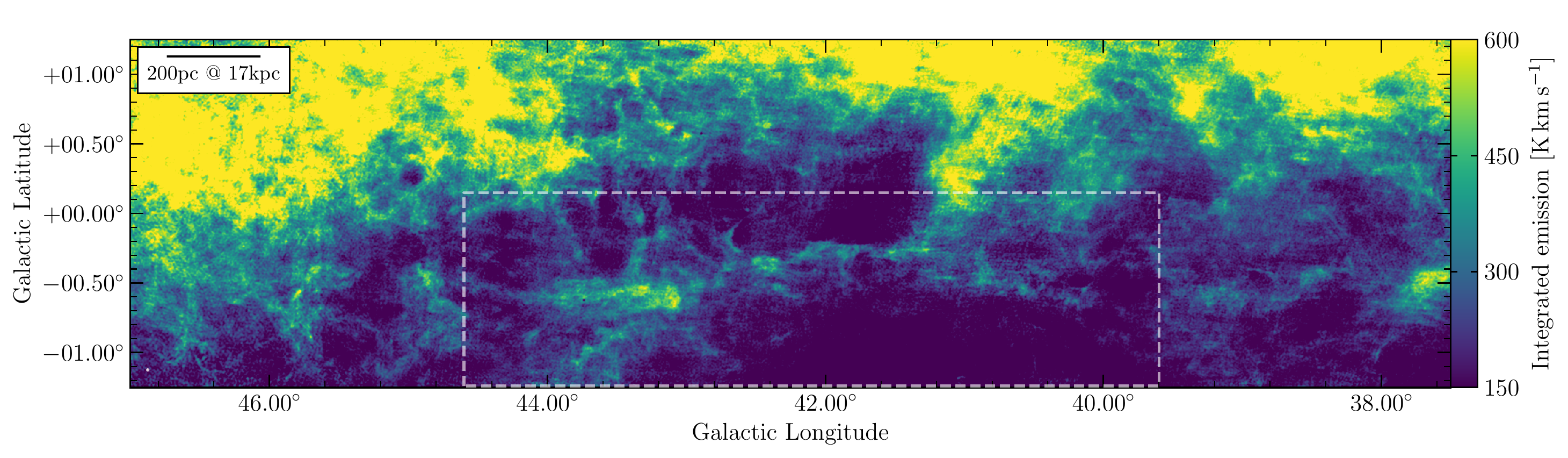}
      \caption[]{Overview of Maggie. The map shows the integrated \ion{H}{i} emission in the velocity interval between $-$57.5 and $-$48.5$\rm\,km\,s^{-1}$. Maggie is located below the Galactic midplane. The white dashed box containing Maggie marks the region that is considered in the subsequent analysis.}
      \label{fig:HI_overview}
   \end{figure*}

\subsection{Gaussian decomposition}
    
    The THOR \ion{H}{i} emission spectra show Gaussian-like structures at negative velocities, with usually three or fewer components superposed. We therefore use the fully automated Gaussian decomposition algorithm \textsc{GaussPy+}\footnote{\url{https://github.com/mriener/gausspyplus}} \citep{2019A&A...628A..78R} to study the kinematics of the filament.
    
    \textsc{GaussPy+} is a multi-component Gaussian decomposition tool based on the earlier \textsc{GaussPy} algorithm \citep{2015AJ....149..138L} and provides an improved fitting routine and a fully automated means to decompose emission spectra using machine-learning algorithms. \textsc{GaussPy+} automatically determines initial guesses for Gaussian fit components using derivative spectroscopy. To decompose the spectra, the spectra require smoothing to remove noise peaks while retaining real signal. The optimal smoothing parameters are found by employing a machine-learning algorithm that is trained on a subsection of the data set.
    
\begin{figure*}
      \centering
        \includegraphics[width=1.0\textwidth]{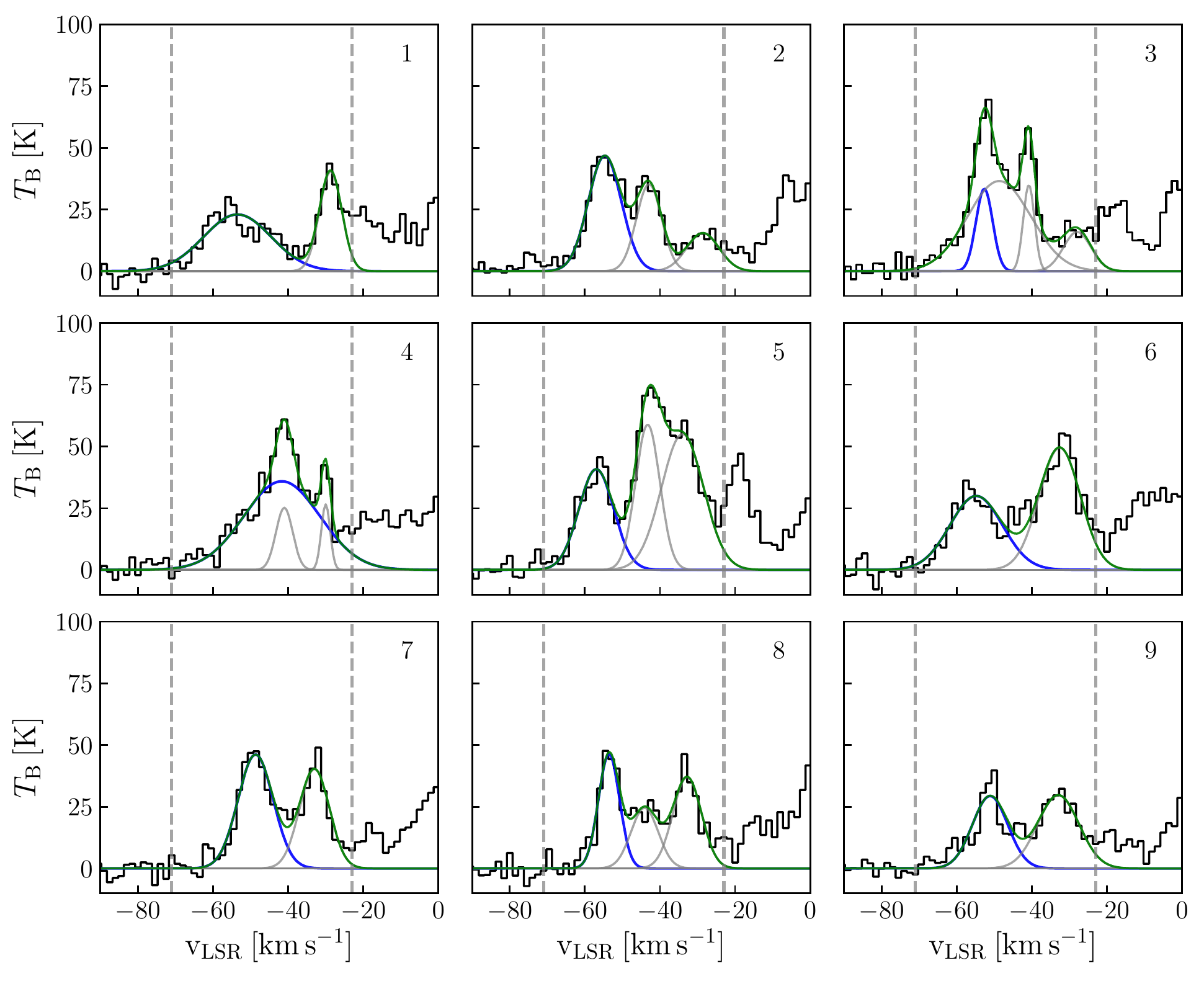}
      \caption[]{\ion{H}{i} spectra and \textsc{GaussPy+} decomposition results along the filament. The black curves show \ion{H}{i} spectra corresponding to the positions marked in the middle panel of Fig.~\ref{fig:Maggie_overview}. The vertical dashed lines at $-71$ and $-23\rm\,km\,s^{-1}$ mark the velocity range taken into account for the \textsc{GaussPy+} decomposition. The blue curve in each panel shows the first (i.e. lowest-centroid-velocity) component, which we attribute to Maggie. The gray curves show all additional components that were fitted. The green curve represents the total spectrum fit in the considered velocity range.}
      \label{fig:gpy_spectra}
    \end{figure*}

    As Maggie is found at negative velocities and the \ion{H}{i} emission toward the inner Galactic plane with $\varv_{\mathrm{LSR}}\gtrsim 0\rm\,km\,s^{-1}$ is ubiquitous in the spectra, it would not be sensible to fit the whole spectra with \textsc{GaussPy+}. Instead, to save computational resources and achieve a better decomposition performance, we masked all spectral channels at velocities $\leq -71\rm\,km\,s^{-1}$ and $\geq -23\rm\,km\,s^{-1}$ (see Fig.~\ref{fig:gpy_spectra}). Maggie has velocities around $-54\rm\,km\,s^{-1}$ (see Fig.~\ref{fig:Maggie_overview} and Sect.~\ref{sec:kinematics}) and we aim to disentangle components that might blend in with Maggie.
    
    However, it is essential to reliably estimate the noise in the spectra to obtain good fit results. \textsc{GaussPy+} comes with an automated noise estimation routine as a preparatory step for the decomposition. For this step, we supplied the full spectra to \textsc{GaussPy+} as the masked spectra might not contain enough noise channels. The noise map derived in that way was then used for the decomposition of Maggie. 
    
    We ran the \textsc{GaussPy+} training step with 500 randomly-selected spectra from the \ion{H}{i} data to find the optimal smoothing parameters for the fitting, as recommended in \citet{2019A&A...628A..78R}. For \ion{H}{i} observations, we would expect both narrow and broad line widths owing to the multiphase (i.e. WNM-CNM-UNM) nature of \ion{H}{i} emission. To account for that, we chose not to refit broad or blended components, parameters that can easily be adjusted in the \textsc{GaussPy+} routine.
    
     After the initial fitting, the algorithm applies a two-phase spatial coherence check that can optimize the fit by refitting the components based on the fit results of neighboring pixels. Figure~\ref{fig:gpy_spectra} shows example spectra along the Maggie filament marked in Fig.~\ref{fig:Maggie_overview}, and the final fit results from the \textsc{GaussPy+} decomposition. Between two and three components were typically fitted by \textsc{GaussPy+} in the given velocity range. In every pixel spectrum, we selected the component with the lowest centroid velocity as the ``Maggie'' component (see Sect.~\ref{sec:kinematics}) since no significant portion of \ion{H}{i} emission is found at $\varv_{\mathrm{LSR}}\lesssim -60\rm\,km\,s^{-1}$. Due to blended components and strong emission at higher velocities, a selection based on amplitude of the components fails to recover the structure identified in \citet{2020A&A...642A.163S} as the Maggie filament.
    
%
   
%-----------------------------------------------------------------

\section{Results}\label{sec:results}

\subsection{Location, distance, and morphology}\label{sec:location}
    Figure~\ref{fig:HI_overview} shows the integrated \ion{H}{i} emission covering the Galactic plane at longitudes and latitudes $37.5^{\circ}<\ell<47.0^{\circ}$ and $\vert b\vert<1.25^{\circ}$, respectively. The filament Maggie subtends an area from $(\ell,b)=(40.4,-0.3)^{\circ}$ to $(44.2,-0.9)^{\circ}$ in Galactic longitude and latitude, and has a velocity around $-54\rm\,km\,s^{-1}$ with respect to the local standard of rest ($\rm LSR$). In this region in position-position-velocity ($p$-$p$-$\varv$) space, the Galactic \ion{H}{i} disk shows a warp toward higher latitudes \citep[see e.g.,][]{1988gera.book..295B,1990ARA&A..28..215D,1993AIPC..278..447S,2009ApJ...693.1250D}, which places Maggie at a location significantly displaced from the Galactic midplane. The \ion{H}{i} midplane lies even beyond the coverage of the THOR survey at $b\geq+1.25^{\circ}$. Therefore, the midplane and Maggie are separated by at least $\Delta b\geq1.6^{\circ}$ on the plane of the sky.
    
    In the forthcoming analysis we assumed the Galactic rotation model by \citet{2019ApJ...885..131R} and used the Bayesian distance calculator of the BeSSeL survey \citep{2016ApJ...823...77R} to translate the location in $p$-$p$-$\varv$ space into a physical distance. Maggie's estimated distance away from us is $d=17\pm1\rm\,kpc$ and its distance to the Galactic center is $R_{\rm GC}=12\pm1\rm\,kpc$. This physical scale locates the filament $\geq470\rm\,pc$ below the Galactic midplane, which is greater than the average \ion{H}{i} scale height ($\sim$200$\rm\,pc$) at this Galactocentric radius \citep{2009ARA&A..47...27K}.
    
    The Maggie filament discloses a hub-like feature in the east, on which smaller-scale filaments appear to converge, and a tail that thins out toward the west. The northwestern part shows a connection to the midplane, potentially feeding off the \ion{H}{i} material located at higher latitudes. Most of the filament, however, appears to be disconnected from the Galactic midplane material.
    
\begin{figure*}
      \centering
        \includegraphics[width=1.0\textwidth]{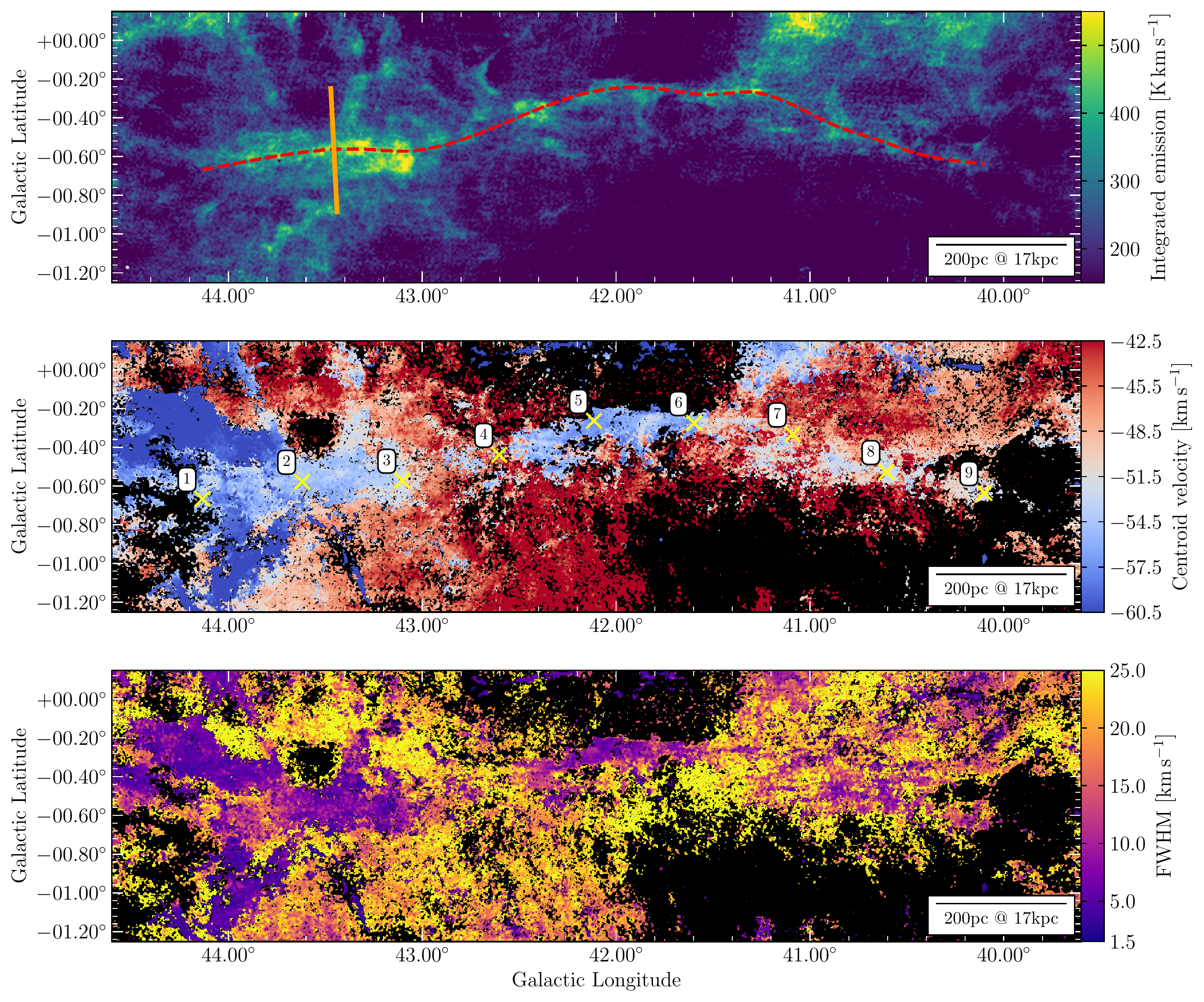}
      \caption[]{Kinematic overview of the Maggie filament. \textit{Top panel:} Integrated \ion{H}{i} emission in the velocity interval between $-$57.5 and $-$48.5$\rm\,km\,s^{-1}$. The red dashed curve marks the spine of the filament. The orange line perpendicular to the spine marks the cut of the average emission profile shown in black in the upper panel of Fig.~\ref{fig:Maggie_width}. \textit{Middle panel:} Centroid velocities of the first (or single) component derived from the \textsc{GaussPy+} decomposition of the emission spectra. The fits are restricted to the velocity range $-71\leq \varv_{\mathrm{LSR}}\leq-23\rm\,km\,s^{-1}$. The yellow crosses mark the positions of the spectra shown in Fig.~\ref{fig:gpy_spectra}. \textit{Bottom panel:} Corresponding line widths in terms of the full width at half maximum (FWHM). Both maps are clipped at $5\sigma$ ($\sim$20$\rm\,K$).}
      \label{fig:Maggie_overview}
   \end{figure*}

    We define a backbone that runs through Maggie by selecting a spline based on visual inspection of the integrated emission (see top panel in Fig.~\ref{fig:Maggie_overview}). The length and width of Maggie are estimated using the Filament Characterization Package\footnote{\url{https://github.com/astrosuri/filchap}} \citep[\textsc{FilChaP};][]{2019A&A...623A.142S}. These properties are not heavily affected by a potential misplacement of Maggie's spine.
    
    The length of the filament as marked by the red spine in the top panel of Fig.~\ref{fig:Maggie_overview} is $\ell_{\mathrm{tot}}=1.2\pm 0.1\rm\,kpc$ assuming a distance of $17\rm\,kpc$. The uncertainty in length is dominated by the uncertainty in distance. The width of the filament was estimated using the integrated emission map shown in Fig.~\ref{fig:Maggie_overview}. We took perpendicular slices of $200\rm\,pc$ width with a step size of $10\rm\,pc$ ($\sim$3$\times$beam) along the filament. We then averaged the radial emission profiles over three neighboring slices and fitted Gaussian and Plummer-like functions to the averaged emission profile.
    Plummer-like functions (as functions of cylindrical radius $r$) have the form
    \begin{equation}
        I(r) \propto \frac{\rho_{\mathrm{c}}R_{\mathrm{flat}}}{\left[1+\left(r/R_{\mathrm{flat}}\right)^2\right]^{\frac{p-1}{2}}} \: ,
    \end{equation}
    \noindent where $p$ is the power-law index, $\rho_{\mathrm{c}}$ is the density at the center of the filament, and $R_{\mathrm{flat}}$ is the characteristic radius of the flat inner portion of the density profile. More details about the fitting can be found in \citet{2019A&A...623A.142S}.
    
    Plummer-like functions are used to describe the radial column density (i.e. integrated emission in the optically thin limit) profile of a cylindrically shaped filament that has a flat density distribution up to $R_{\mathrm{flat}}$ and a power-law fall-off beyond \citep{2011A&A...529L...6A}.
    We used constant power-law indices $p=2$ and 4 for the Plummer-like functions due to the degeneracy between $\rho_{\mathrm{c}}$ and $R_{\mathrm{flat}}$ (and therefore $p$) in finding the best fit solution \citep[see][]{2014MNRAS.445.2900S}. 
    The power law index $p=4$ describes the \citet{1964ApJ...140.1056O} model of an isothermal filament in hydrostatic equilibrium. The power-law index $p=2$ is derived by \citet{2011A&A...529L...6A} as a free fit parameter from small-scale filamentary structures seen in \textit{Herschel} data and might be attributed to non-isothermal or magnetized filaments \citep[see][and references therein]{2011A&A...529L...6A}.
    
\begin{figure}
      \centering
        \resizebox{\hsize}{!}{\includegraphics{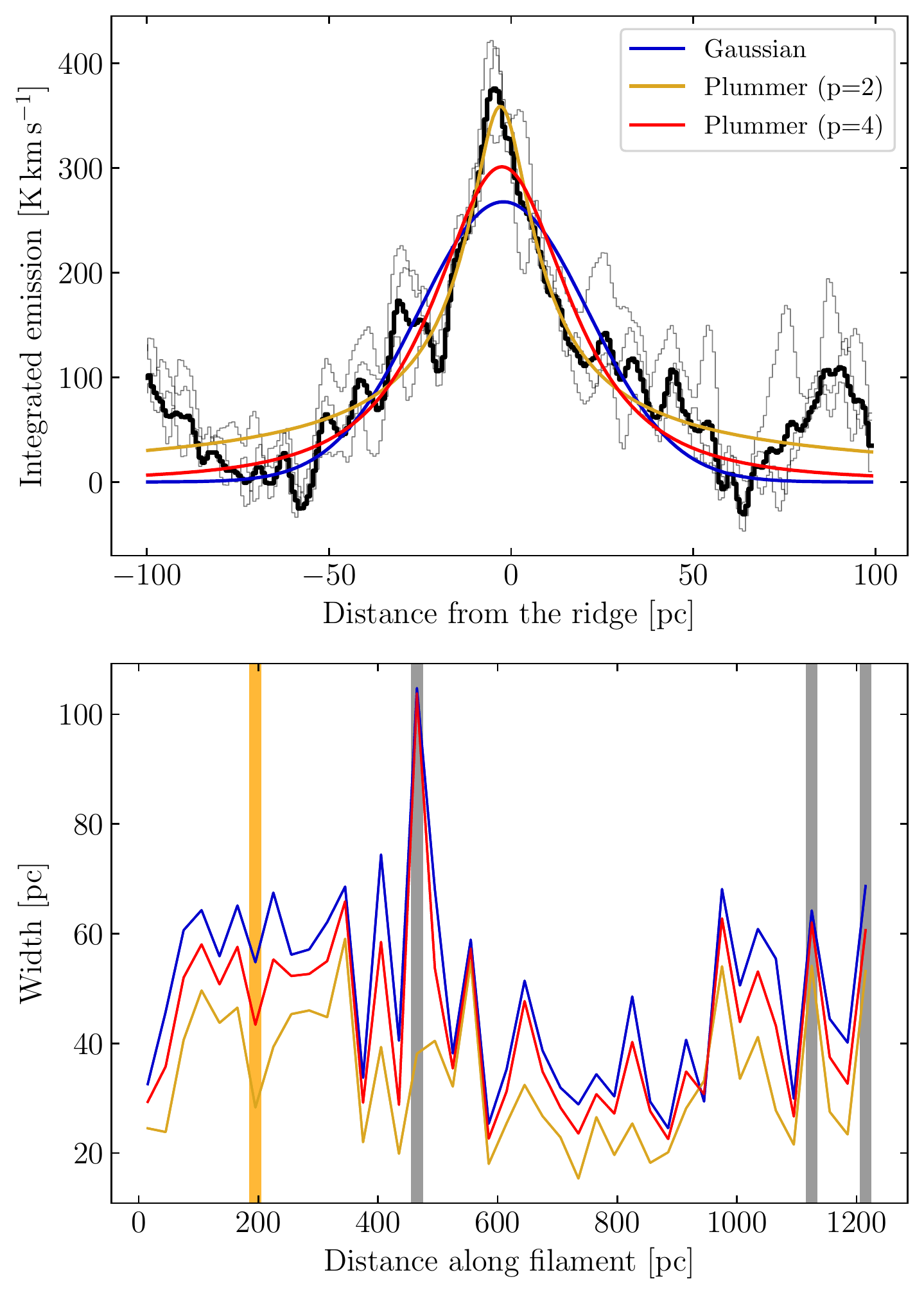}}
      \caption[]{Width of the Maggie filament. \textit{Top panel:} Example emission profile perpendicular to the filament. At each step along the filament, three neighboring emission profiles (gray) are taken to estimate an average profile (black). A fitted baseline has been subtracted from each profile. The cut along which the average emission profile is measured is marked in the upper panel of Fig.~\ref{fig:Maggie_overview}. To derive the widths, we fitted the average emission profiles with Gaussian (blue), and Plummer-like functions with index $p=2$ (yellow) and $p=4$ (red), respectively. \textit{Bottom panel:} The fitted widths of Maggie are plotted against the distance along the filament. The colors correspond to the fits shown in the upper panel. The gray bars show the widths that are discarded as these profiles could not be fitted properly. The orange bar marks the position of the example profile shown in the upper panel.}
      \label{fig:Maggie_width}
   \end{figure}

    The fitted widths along the filament are shown in Fig.~\ref{fig:Maggie_width}, with zero distance at the outermost point in the east. The widths derived from Gaussian fits are systematically larger than the widths estimated by Plummer-like functions. While the Plummer-like function with an index of $p=2$ is able to reliably reproduce individual peaks (upper panel of Fig.~\ref{fig:Maggie_width}), the fits on average underestimate the width of the emission profiles. Gaussian fits do not fully recover the amplitude of the profiles but give a more accurate representation of the wings. As the noise and blending of different emission features make it difficult to determine which function represents the emission profiles best, we consider the fits equally good. The reduced chi-squared distributions of the fits are similar and do not allow any preference of a fitting function. Three positions along the filament (vertical gray bars in Fig.~\ref{fig:Maggie_width}) are discarded as the fits gave no reliable results, regardless of the fitting function.
    For simplicity, we report the average width taken over the full length of the filament. Taking into account all fitting methods, the average width is $w=40\pm6\rm\,pc$, which is well resolved by the beam of the THOR observations corresponding to a spatial scale of $\approx 3.3\rm\,pc$. The uncertainty is dominated by the variance of the measured widths along the filament. The filament width is only weakly affected by the velocity integration range. If we integrate the emission over the velocity range --57.5 to --42.5$\rm\,km\,s^{-1}$, the average filament width becomes 44\,pc, which is well within the uncertainty of our derived width. However, due to blending emission components toward higher velocities, the derivation of the filament width becomes increasingly difficult with increasing velocity integration range. We therefore constrain the integration range to roughly the lower and upper quartile of the velocity distribution (see Sect.~\ref{sec:kinematics}).
    
    One of the characteristics that is usually used to define filaments is the projected aspect ratio \citep[see e.g.,][]{2015ApJ...815...23Z,2018ApJ...864..153Z}. Taking the average width, Maggie has an aspect ratio of (30:1). Although Maggie is identified through atomic line emission, as opposed to the selection of molecular filaments investigated by \citet{2018ApJ...864..153Z}, Maggie roughly agrees with the overall aspect ratios of other large-scale filaments \citep{2015MNRAS.450.4043W} and the identified Milky Way bones \citep{2014ApJ...797...53G,2015ApJ...815...23Z}. 
    
\subsection{Kinematics}\label{sec:kinematics}
   
   In the following section, we focus in detail on the kinematic properties of Maggie. We show in the middle and bottom panel of Fig.~\ref{fig:Maggie_overview} the maps of the fitted centroid velocities and line widths of the first fit component (corresponding to the lowest-velocity component of the \textsc{GaussPy+} decomposition), to which we refer as the ``Maggie component'' in the following. Accordingly, the second fit component refers to the next component at higher centroid velocity. Overall, the peak velocities along the filament vary between $\sim -60$ and $\sim -45\rm\,km\,s^{-1}$ and exhibit an undulating pattern (see also Fig.~\ref{fig:pv_filament}). For the most part, the velocity map clearly reflects the spine of the filament, as components off the spine have significantly larger velocities but are picked up as the first component in the absence of Maggie. Given the large spatial scales, the velocities along the spine show a smooth gradient of less than $\pm 3\rm\,km\,s^{-1}\,(10\,pc)^{-1}$, a kinematic property that satisfies the bone criteria in \citet{2015ApJ...815...23Z}.
   
   \begin{figure}
      \centering
        \resizebox{\hsize}{!}{\includegraphics{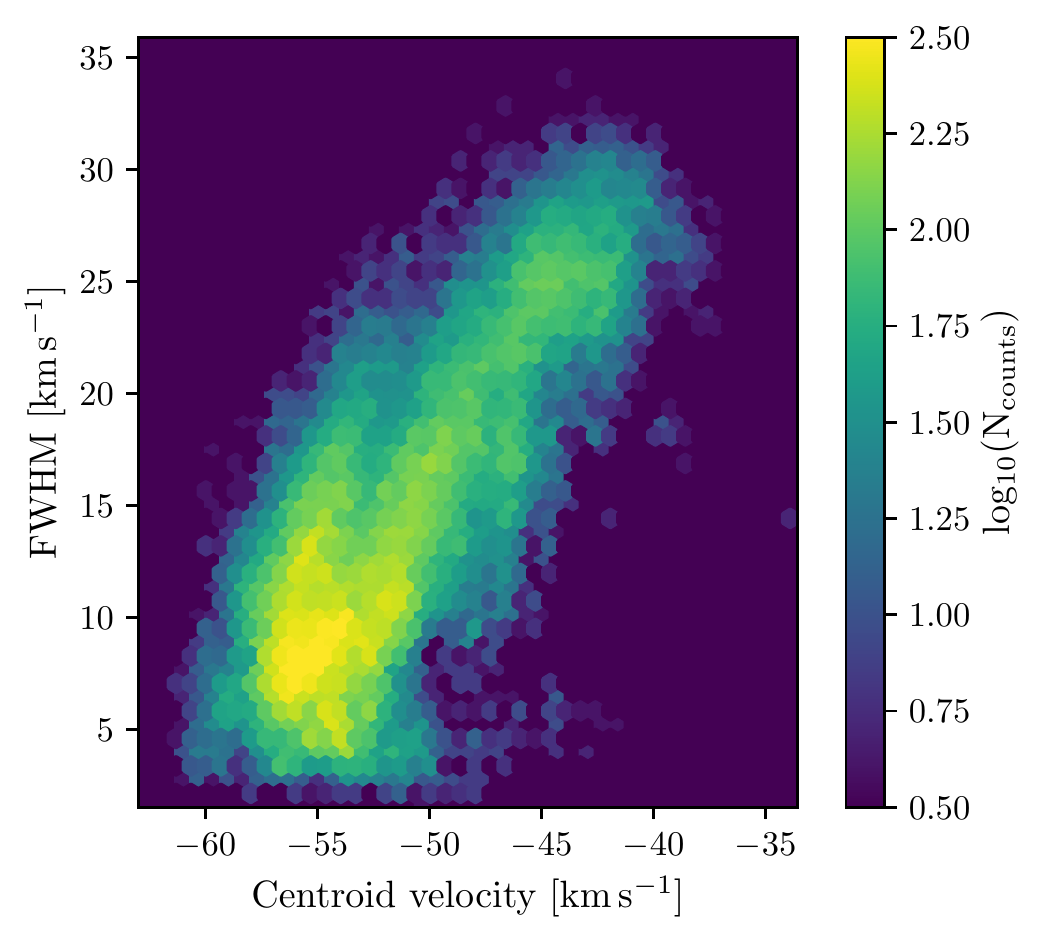}}
      \caption[]{Line width--centroid velocity correlation. The scatter plot shows the line width (FWHM) as a function of the centroid velocity at all pixel positions along the 40\,pc-wide spine of Maggie.}
      \label{fig:fwhm_velo_corr}
   \end{figure}
   
   The line widths as well as centroid velocities shown in Fig.~\ref{fig:Maggie_overview} indicate a stark contrast between the filament and the background. Furthermore, the kinematics in terms of the centroid velocity and line width along the filament spine are moderately correlated, where narrow line width corresponds to lower centroid velocity. Figure~\ref{fig:fwhm_velo_corr} presents the relationship between the centroid velocity and line width along the 40\,pc-wide spine of the filament. The Pearson correlation coefficient between the velocity and line width distribution is 0.76, implying that under the assumption of a linear dependence $\sim$60\% of the variance in the line width distribution can be explained by the scatter present in the centroid velocity distribution. We note that this does not necessarily imply a causal relation between the line width and velocity. Rather, this correlation is likely to be due to contamination from background gas at positions where the filament emission is less well defined and difficult to disentangle from blending components.
   
   \begin{figure}
      \centering
        \resizebox{\hsize}{!}{\includegraphics{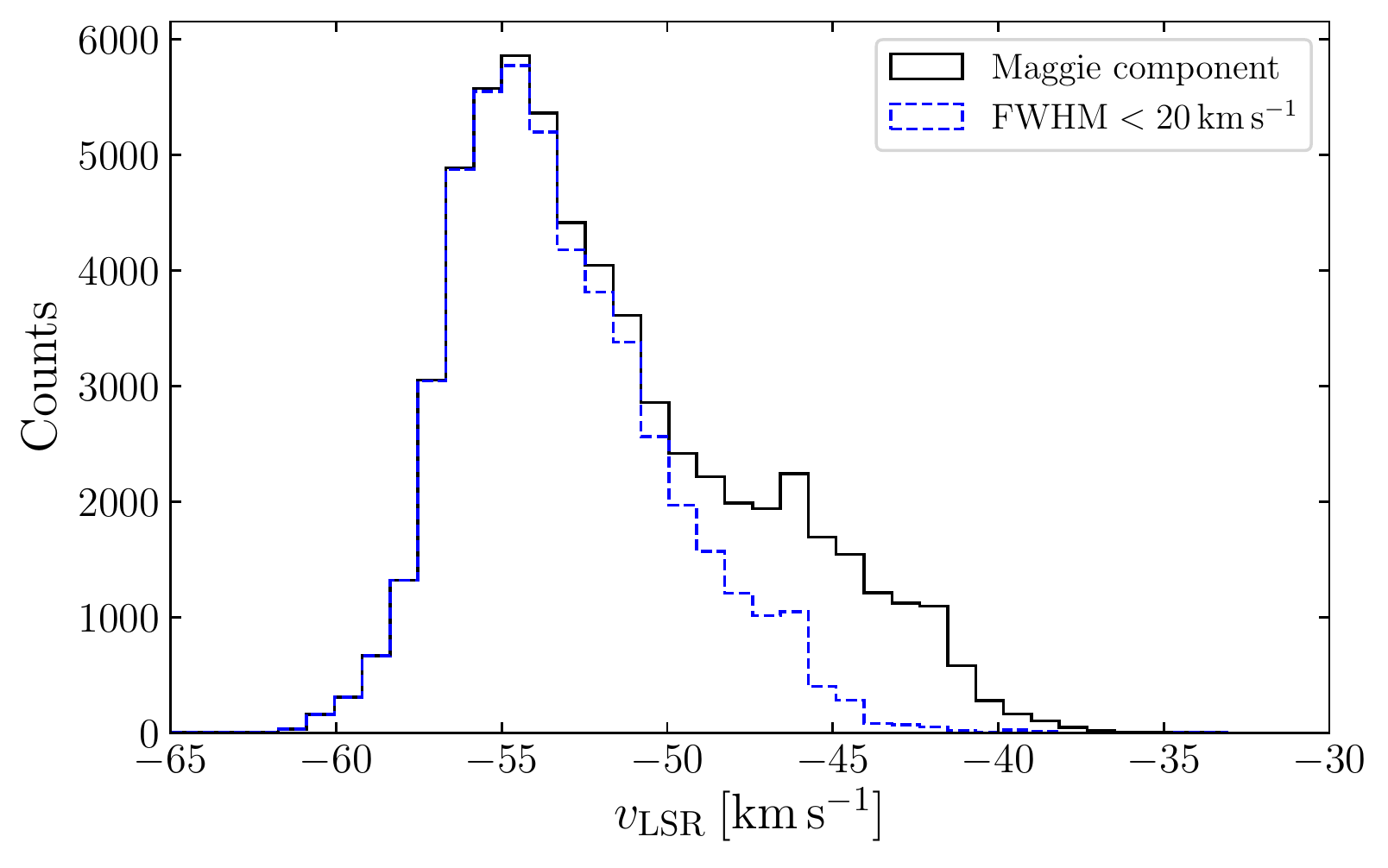}}
      \caption[]{Histograms of fitted peak velocities of Maggie. The black histogram shows the distribution of the centroid velocities at all pixel positions inferred from the \textsc{GaussPy+} fits along the $40\rm\,pc$-wide spine of the filament. The blue histogram shows the peak velocity distribution corresponding to components with line widths $\Delta \varv < 20\rm\,km\,s^{-1}$.}
      \label{fig:hist_velo}
   \end{figure}
   
   Figure~\ref{fig:hist_velo} shows the histogram of centroid velocities of Maggie, where all individual pixel positions along the $40\rm\,pc$-wide spine of the filament are being sampled. The centroid velocities along the filament exhibit a strong peak at $-54\rm\,km\,s^{-1}$, with some skewness toward $-45\rm\,km\,s^{-1}$ that might be attributed to velocity blending of broad WNM components at higher velocities.
   
    \begin{figure}
      \centering
        \resizebox{\hsize}{!}{\includegraphics{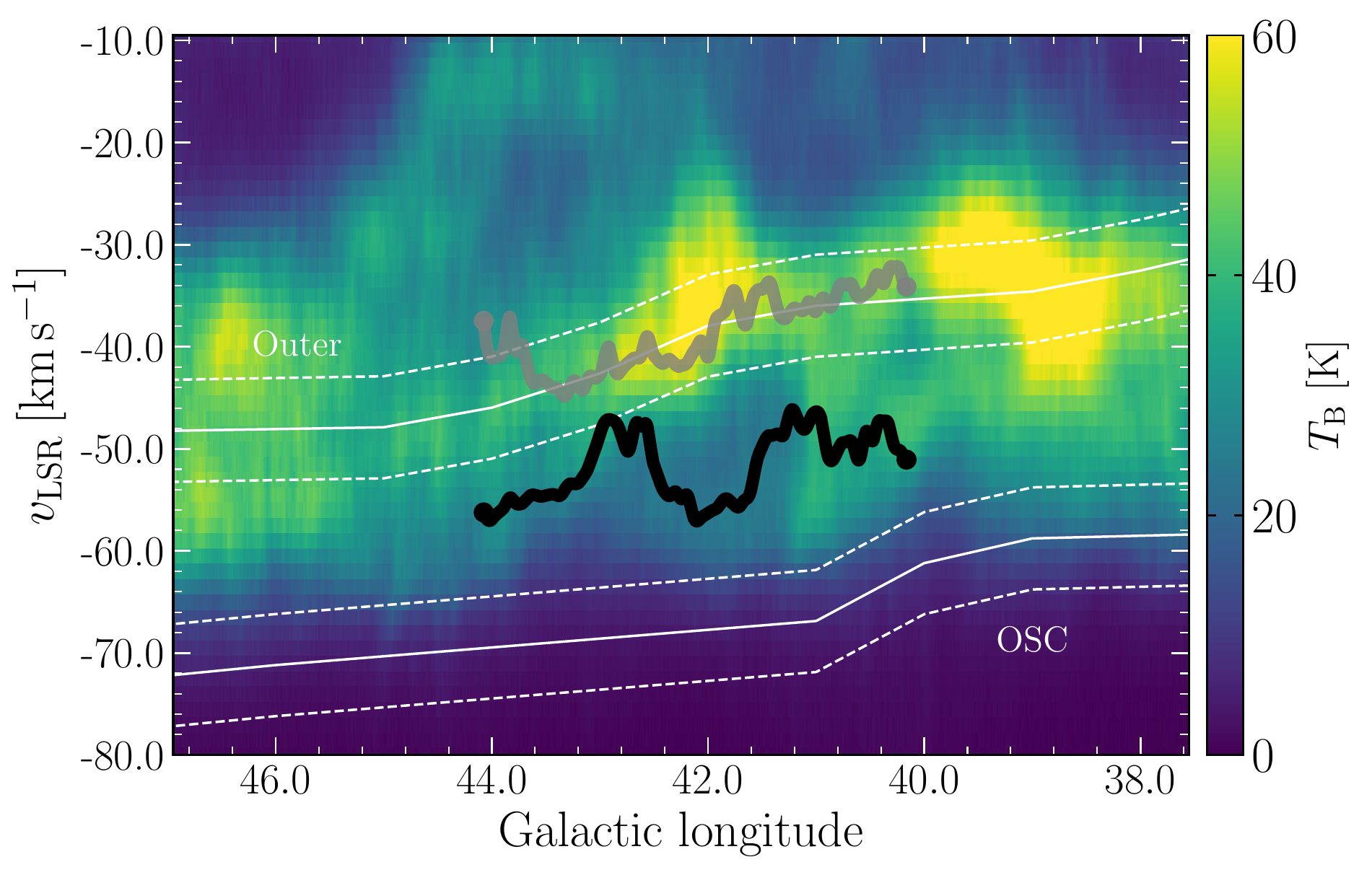}}
      \caption[]{Position-velocity diagram of \ion{H}{i}. The colored background shows the \ion{H}{i} emission averaged over $-1.25^{\circ} < b < +1.25^{\circ}$. The black and gray curve indicate the intensity-weighted average velocity of Maggie and the second fit component at each longitude as given by the \textsc{GaussPy+} decomposition, respectively. The white line segments give the predicted spiral arm centroids of the Outer arm and Outer-Scutum-Centaurus (OSC) arm along with their uncertainties (dashed lines) in velocity \citep{2019ApJ...885..131R}.}
      \label{fig:Maggie_pv_plot}
   \end{figure}
   
   Figure~\ref{fig:Maggie_pv_plot} offers a position-velocity view of Maggie and its location with respect to the average \ion{H}{i} emission. We also overplot the predicted locations of the spiral arms as given by \citet{2019ApJ...885..131R}. The bulk of \ion{H}{i} material and the second fit component of our spectral decomposition trace out the spine of the Outer arm remarkably well while Maggie exhibits an offset of 5--15$\rm\,km\,s^{-1}$. Given the distance from the midplane, neither spatially nor kinematically can Maggie be associated with any spiral arm structure. We note, however, that different spiral arm models can vary by $10\rm\,km\,s^{-1}$ or more in position-velocity space, a concern raised by \citet{2015ApJ...815...23Z} when classifying bones of the Milky Way.
   
\begin{figure}
      \centering
        \resizebox{\hsize}{!}{\includegraphics{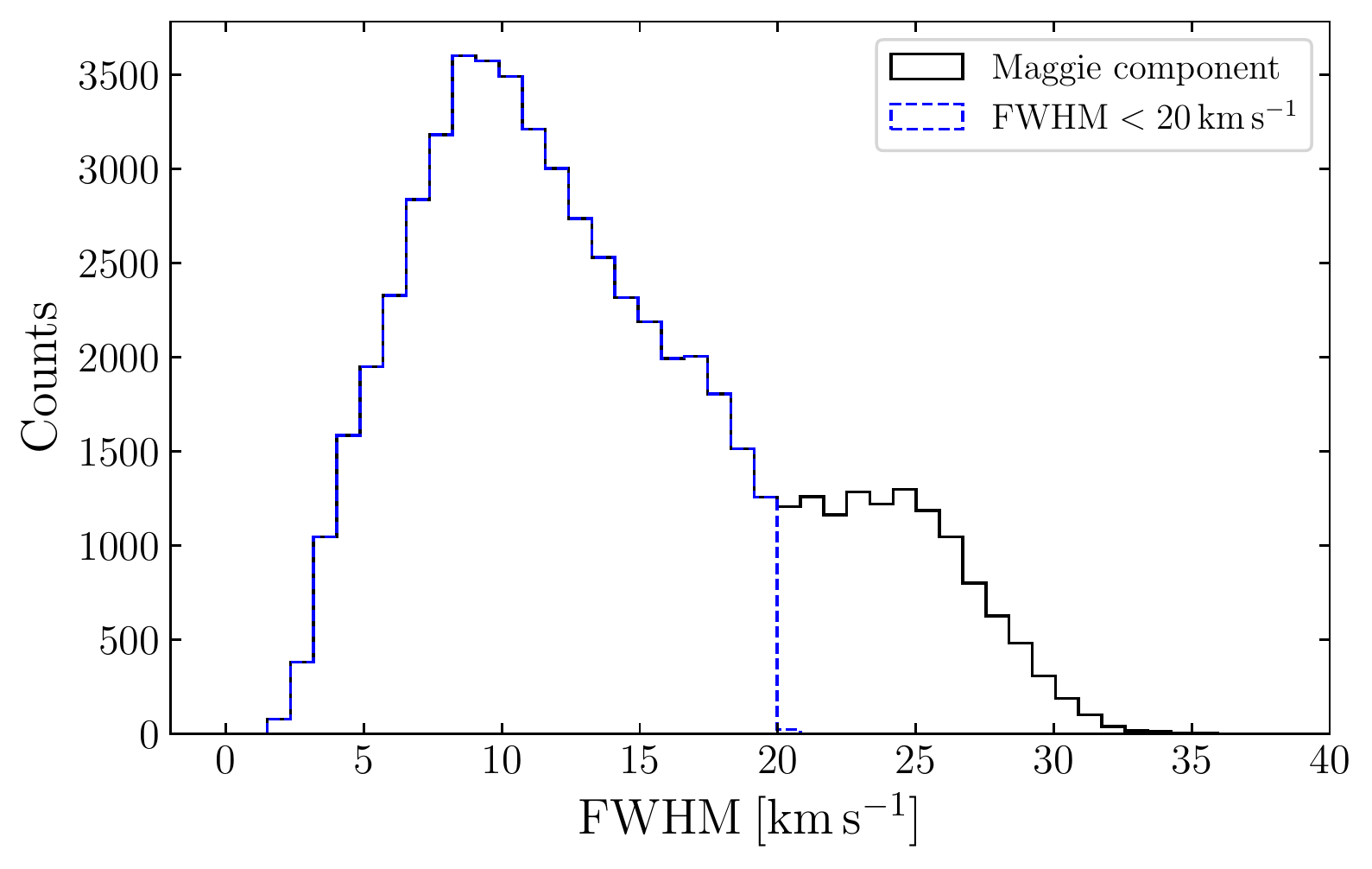}}
      \caption[]{Histograms of fitted line widths of Maggie. The black histogram shows the distribution of the line widths at all pixel positions in terms of the FWHM inferred from the \textsc{GaussPy+} fits along the $40\rm\,pc$-wide spine of the filament. The blue histogram shows the line width distribution corresponding to components with line widths $\Delta \varv < 20\rm\,km\,s^{-1}$.}
      \label{fig:hist_fwhm}
   \end{figure}

\begin{figure}
      \centering
        \resizebox{\hsize}{!}{\includegraphics{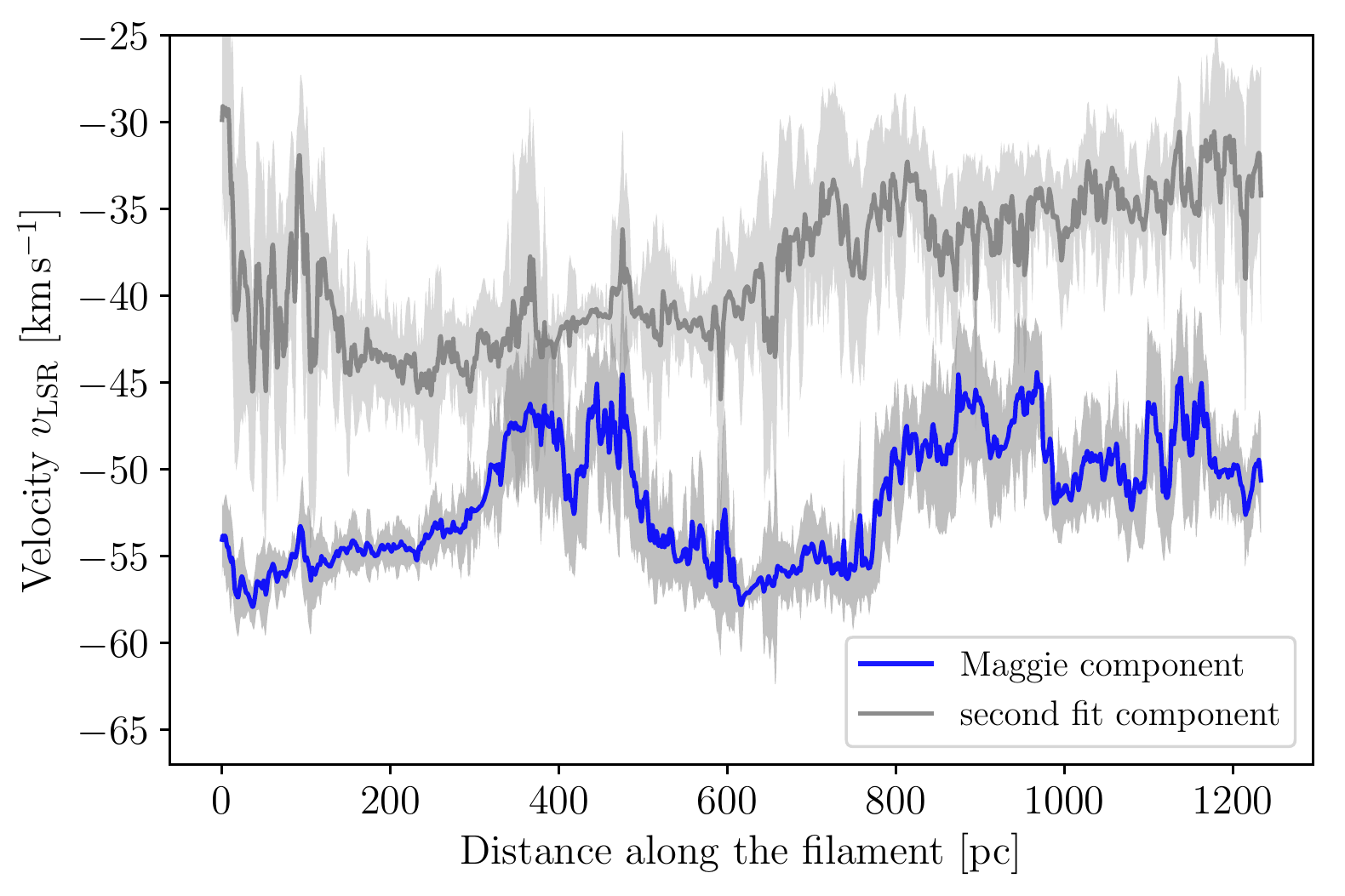}}
      \caption[]{Position-velocity plot along the filament. This plot shows the centroid velocity over distance along the filament marked by the red spine in the upper panel of Fig.~\ref{fig:Maggie_overview}. The centroid velocity is an intensity-weighted average over the $40\rm\,pc$-wide slice perpendicular to every filament position. The blue curve represents the designated Maggie component, that is the first fit component of the \textsc{GaussPy+} results. The gray curve shows the second fit component for comparison. The gray shaded area around both curves indicates the uncertainty estimated by the standard deviation over each slice.}
      \label{fig:pv_filament}
   \end{figure}

   The histogram of line widths in Fig.~\ref{fig:hist_fwhm} shows a broad distribution, with line widths between $4$ and $30\rm\,km\,s^{-1}$. The histogram has a peak around $10\rm\,km\,s^{-1}$ and a shoulder that extends to $25\rm\,km\,s^{-1}$.
   We additionally show in Figs.~\ref{fig:hist_velo} and \ref{fig:hist_fwhm} the histograms corresponding to velocity components with a line width $\Delta \varv < 20\rm\,km\,s^{-1}$, discarding the shoulder in the line width distribution. The median values of the total centroid velocity distribution is $-52.7\rm\,km\,s^{-1}$, and only taking into account line widths $\Delta \varv < 20\rm\,km\,s^{-1}$, the median shifts to $-53.8\rm\,km\,s^{-1}$.
   
   To study the kinematic structure with respect to the position along the filament we show in Fig.~\ref{fig:pv_filament} the average velocity of each $40\rm\,pc$-wide slice weighted by the amplitude at each pixel position. The uncertainties are estimated by the standard deviation of each slice. The wave-like velocity structure of Maggie is now more clearly visible. The Maggie velocities beyond $>-48.5\rm\,km\,s^{-1}$ around $400\rm\,pc$ distance (corresponding to the region around spectrum 4 in Fig.~\ref{fig:gpy_spectra}) are difficult to clearly separate from the second fit component. We note that the mean velocity and particularly line width have to be treated with some caution as the velocity blending makes it difficult to distinctively identify Maggie in this part of the region. We revisit the velocity structure in Sect.~\ref{sec:structure_function} to investigate signatures of characteristic spatial scales.
   
\subsection{Column density and mass}\label{sec:coldens_mass}

    Measuring \ion{H}{i} absorption against strong continuum sources allows us a direct derivation of the optical depth. We therefore use the high angular resolution \ion{H}{i}+continuum data, that consist of VLA C-array observations only, to filter out large-scale emission and measure \ion{H}{i} absorption against discrete continuum sources. We measured HI+continuum spectra against seven continuum sources taken from \citet{2018A&A...619A.124W} that have brightness temperatures $\geq 200\rm\,K$ for a sufficient signal-to-noise ratio. The sources are located both within and outside the area of Maggie (i.e. within and outside the contour at $4.2\,M_{\odot}\rm\,pc^{-2}$, see later in this section). Since the synthesized beam of the native C-configuration data is $\sim$14$\arcs$, we extracted an average \ion{H}{i}+continuum spectrum from a $16\arcs\times 16\arcs$ ($4\times 4$ pixels) area. To the first order, we can then estimate the \ion{H}{i} optical depth by \citep[see][]{2015A&A...580A.112B}
    \begin{equation}
        \tau_{\rm{simplified}} = -\mathrm{ln}\left(\frac{T_{\rm{on}}}{T_{\rm{cont}}}\right) \: ,
        \label{equ:HICA_tau_simplified}
    \end{equation}
    \noindent where $T_{\rm{on}}$ is the \ion{H}{i} brightness temperature against the continuum source and $T_{\rm{cont}}$ is the brightness of the continuum source. Using the optical depth, we compute the \ion{H}{i} spin temperature with
    \begin{equation}
        T_{\mathrm{s}} = \frac{T_{\rm{off}}}{1-e^{-\tau}} \: .
        \label{equ:T_spin}
    \end{equation}
    \noindent We measured the \ion{H}{i} brightness at an offset position $T_{\rm{off}}$ using the combined THOR \ion{H}{i} data (see Sect.~\ref{sec:observation}). We therefore selected an annulus around each source with inner and outer radii of 60\arcsec and 120\arcsec, respectively, to measure an averaged $T_{\rm{off}}$.
    
    All spectra measured against Galactic \ion{H}{ii} regions do not exhibit any absorption features in the velocity range of Maggie (Tab.~\ref{tab:continuum_sources}). This gives additional support to the argument that the Maggie filament is in fact on the far side of the Galaxy as Galactic \ion{H}{ii} regions are located in the foreground and can therefore not be detected in absorption. Furthermore, we find that no absorption is evident in the spectra of both Galactic and extragalactic continuum sources offset from the line of sight of Maggie.
    
    \begin{table*}
        \caption{Continuum sources.}
        \renewcommand*{\arraystretch}{1.3}
        \centering
        \begin{tabular}{c c c c c c c}
             \hline\hline
             Source ID & Glon [$^{\circ}$] & Glat [$^{\circ}$] & $T_{\mathrm{cont}}$ [K] & Type & L.o.s. & Absorption \\\hline
             G43.921-0.479 & 43.92 & -0.48 & 304 & extragal. & \cmark & \cmark \\
             G43.890-0.783 & 43.89 & -0.78 & 481 & \ion{H}{ii} & \xmark & \xmark \\
             G43.738-0.620 & 43.74 & -0.62 & 503 & extragal. & \cmark & \cmark \\
             G43.177-0.519 & 43.18 & -0.52 & 296 & \ion{H}{ii} & \cmark & \xmark \\
             G42.434-0.260 & 42.43 & -0.26 & 277 & \ion{H}{ii} & \xmark & \xmark \\
             G42.028-0.605 & 42.03 & -0.60 & 519 & extragal. & \xmark & \xmark \\
             G41.513-0.141 & 41.51 & -0.14 & 205 & \ion{H}{ii} & \xmark & \xmark \\
        \end{tabular}
        \tablefoot{\textit{Column} 1 gives the source name as listed in the continuum catalog by \citet{2018A&A...619A.124W}. \textit{Columns} 2 and 3 are the Galactic longitude and latitude, respectively. \textit{Column} 4 denotes the average brightness temperature of the continuum source over an area of $16\arcs\times 16\arcs$. \textit{Column} 5 describes the physical nature of the continuum source. \textit{Column} 6 marks if the continuum source is located along the line of sight (L.o.s), that is within the contour at $4.2\,M_{\odot}\rm\,pc^{-2}$ (see Sect.~\ref{sec:coldens_mass}). \textit{Column} 7 indicates if absorption above $3\sigma$ is detected in the velocity range of Maggie.}
        \label{tab:continuum_sources}
    \end{table*}{}
    
\begin{figure}
      \centering
        \resizebox{\hsize}{!}{\includegraphics{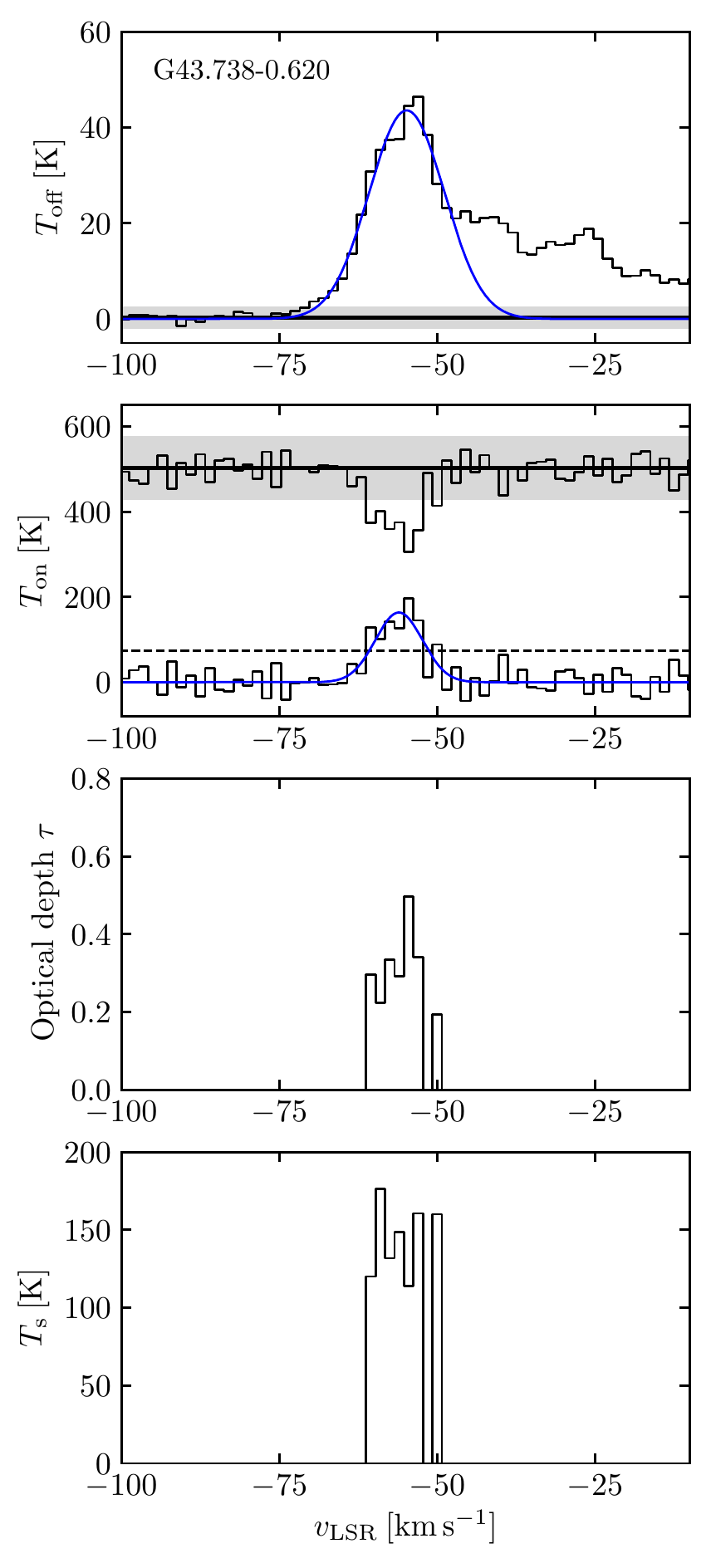}}
      \caption[]{HI emission and absorption spectrum of the extragalactic continuum source G43.738-0.620. The \textit{top panel} shows the emission spectrum and is measured in an annulus around the source with inner and outer radii of $60\arcsec$ and $120\arcsec$, respectively (corresponding to $3-6$ beams of the \ion{H}{i} data). The \textit{second panel} presents the absorption spectrum toward the point source. The lower spectrum indicates the inverted absorption spectrum along with the $3\sigma$ noise (dashed line). In the first two panels, the gray shaded area and blue curve indicate the $3\sigma$ noise and a single-component Gaussian fit, respectively. The \textit{third panel} shows the optical depth computed using Eq.~\eqref{equ:HICA_tau_simplified}. In the \textit{bottom panel} we present the spin temperature, which is computed using  Eq.~\eqref{equ:T_spin}.}
      \label{fig:T_spin_tau}
    \end{figure}

    The extragalactic continuum sources G43.738-0.620 and G43.921-0.479\footnote{The optical depth measurement against G43.921-0.479 is close to the detection limit and will not be discussed further here. However, the mean optical depth $\tau_{\mathrm{mean}}=0.30$ and velocity-integrated optical depth $\int\tau(v)\mathrm{d}v=2.2\rm\,km\,s^{-1}$ are consistent with the other source G43.738-0.620.} are the only sources along Maggie against which we detect \ion{H}{i} in absorption. We show in Fig.~\ref{fig:T_spin_tau} an overview of the optical depth and spin temperature measurement toward G43.738-0.620.
    
    The top panel illustrates the averaged \ion{H}{i} emission profile. In pressure equilibrium, the two stable phases of atomic hydrogen (CNM and WNM) both contribute to the emission. It is difficult to disentangle the properties of either phase from the emission alone, so the information from the absorption profile is critical to learn about the nature of the atomic gas. As the optical depth is proportional to $T_{\mathrm{s}}^{-1}$, we assume that any absorption measured toward the continuum is due to the CNM.
    
    For simplicity we fit both the emission and absorption profile in this specific case (Fig.~\ref{fig:T_spin_tau}) with a single Gaussian component. The line width of the emission profile is $14.1\rm\,km\,s^{-1}$ and therefore significantly wider than the absorption line ($8.8\rm\,km\,s^{-1}$), suggesting that there is contribution to the emission from a broader component that is not evident in the absorption spectrum. However, the line width contributions are not able to pin down the column density fractions of either the CNM or WNM as the (kinetic) gas temperatures are unknown and the line widths can be significantly broadened by non-thermal effects. The total line width, however, can therefore set an upper limit on the kinetic temperature \citep{2003ApJ...586.1067H}. We note that the non-thermal contribution to the CNM component, due to effects such as turbulent motion, should be dominant as the thermal line width of the CNM even with the loose constraint $T_{\mathrm{k}}<500\rm\,K$ \citep[e.g.][]{2001ApJ...551L.105H,2003ApJ...586.1067H} would account for a line width $<4.8\rm\,km\,s^{-1}$.
    
    The optical depth above the $3\sigma$ limit, that we ascribe to the CNM, is plotted in the third panel of Fig.~\ref{fig:T_spin_tau}. The mean and velocity-integrated optical depth over the velocity range of Maggie is $\tau_{\mathrm{mean}}=0.33$ and $\int\tau(v)\mathrm{d}v=2.5\rm\,km\,s^{-1}$, respectively. We note that the reported values in \citet{2020A&A...634A..83W} are inferred from integrating over the whole velocity range of the THOR survey and are thus significantly larger. We use the optical depth and brightness temperature from the emission profile to compute the spin temperature (Eq.~\eqref{equ:T_spin}) in the bottom panel of Fig.~\ref{fig:T_spin_tau}. The mean spin temperature is $\langle T_{\mathrm{s}} \rangle=140\rm\,K$. Owing to the mixing of warm and cold HI along the line of sight, this spin temperature should not be interpreted as a gas kinetic temperature. Rather, it is a density-weighted harmonic mean of the kinetic temperature under the assumption that $T_{\mathrm{s}}=T_{\mathrm{k}}$ \citep[see][]{2000ApJ...536..756D,2003ApJ...585..801D,2009ApJ...693.1250D}. If we assume that the warm HI does not contribute significantly to the optical depth, that the correction for HI self-absorption is negligible and that the CNM has a roughly constant temperature, then we can derive the CNM fraction via
    \begin{equation}
     f_{\mathrm{CNM}} \equiv \frac{N_{\mathrm{CNM}}}{N_{\mathrm{CNM}} + N_{\mathrm{WNM}}} \simeq \frac{T_{\mathrm{CNM}}}{\langle T_{\mathrm{s}}\rangle} \: ,
    \end{equation}
    \noindent where $N_{\mathrm{CNM}}$ and $N_{\mathrm{WNM}}$ are the column density of the CNM and WNM, respectively, and $T_{\rm CNM}$ is the CNM spin temperature. A representative value for the CNM spin temperature should lie within $40 < T_{\rm CNM} < 100\rm\,K$ \citep{2003ApJ...586.1067H,2004ApJ...603..560S,2009ApJ...693.1250D}, which is also in good agreement with the theoretical models by \citet{2003ApJ...587..278W} at the Galactocentric distance $R_{\mathrm{GC}}=11\rm\,kpc$. This conservative estimate gives a CNM fraction in the range $30\% < f_{\rm CNM} < 70\%$. Applying this finding to the whole cloud, Maggie should consequently be composed of a significant fraction of CNM.
    
    We estimate the column density of the atomic hydrogen using \citep[e.g.][]{2013tra..book.....W}
    \begin{equation}
        N_{\ion{H}{i}} = 1.8224\times 10^{18}\int T_{\mathrm{s}}(v)\,\tau(v)\,\mathrm{d}v \: ,
    \end{equation}
    \noindent where $N_{\ion{H}{i}}$ is the total column density of \ion{H}{i} in units of $\rm cm^{-2}$ as a function of the \ion{H}{i} spin temperature $T_{\mathrm{s}}(v)$ and optical depth $\tau(v)$ integrated over the velocity $v$. The spin temperature is $T_{\mathrm{s}}(v)=T_{\mathrm{B}}/(1-e^{-\tau(v)})$, where $T_{\mathrm{B}}$ is the brightness temperature of the \ion{H}{i} emission. The optical depth correction of the column density scales as $\tau/[1-\mathrm{exp}(-\tau)]$, resulting in a correction factor of $1.17$ for $\tau_{\rm mean}=0.33$. We integrated the column density from $-57.5$ to $-48.5\rm\,km\,s^{-1}$, taking into account the predominant velocities of Maggie. The column density then has values in the range $3-10\times 10^{20}\rm\,cm^{-2}$. We used the observed THOR-\ion{H}{i} emission data instead of the individual Maggie component of the decomposition to compute the column density although there might be additional contributions from components blending in at a similar velocity. We show in Appendix~\ref{sec:coldens_app} that the integrated Maggie component is not spatially coherent in amplitude and shows patches of lower amplitude, likely owing to the existence of blended components that are difficult for \textsc{GaussPy+} to disentangle. To estimate a maximum fraction of unrelated blended emission contributing to the column density, we compare the two column density maps. We find that the column density (and ultimately mass) inferred from the single Maggie component (App.~\ref{sec:coldens_app}) is smaller by approximately 30\%.
    
    We note that choosing a single fit component as the ``filament'' can be problematic as the multiphase nature of \ion{H}{i} emission may have multiple components at similar velocities contributing to Maggie. Here, we define Maggie by the centroid velocity of a single component and do not impose any restrictions on the \ion{H}{i} phase of Maggie, which is reasonable in order to study the velocity structure. Yet, by choosing a single component we might include only one \ion{H}{i} phase. \textsc{GaussPy+} does not explicitly check for spatial coherence in amplitude or line width but for coherence in number of components and their centroid velocity. As we argue above, in a CNM-WNM mixture of the gas it is then difficult to characterize Maggie based on the centroid velocity of a single component alone.
    One way to fit and identify multiple phases in \ion{H}{i} emission spectra, is to introduce spatial regularization terms to the loss function of the fitting that are able to cluster different phases (here: CNM and WNM) even if close in velocity \citep{2019A&A...626A.101M}. This approach, however, is computationally expensive and will not be followed here.
    
    The picture of a well-defined CNM phase layer within a WNM envelope assuming thermal pressure equilibrium might be overly simplified. Instead, we would expect a clumpy CNM embedded in a `sea' of WNM, implying that the CNM has a lower volume and area filling factor than the WNM and a strong density contrast \citep{2009ARA&A..47...27K}. The patchy structure seen in Fig.~\ref{fig:coldens_Maggiecomp} could be a reflection of the clumpy structure of the CNM. Lower intensity in small-scale structures could also hint at CNM that has already converted to molecular gas (see Sect.~\ref{sec:atomictransition}). If we want to infer the global multiphase column density of Maggie, however, it is beneficial to examine the properties of the filament by means of the integrated emission rather than using a single \textsc{GaussPy+} fit component, in order to avoid missing mass.
    
\begin{figure*}
      \centering
        \includegraphics[width=1.0\textwidth]{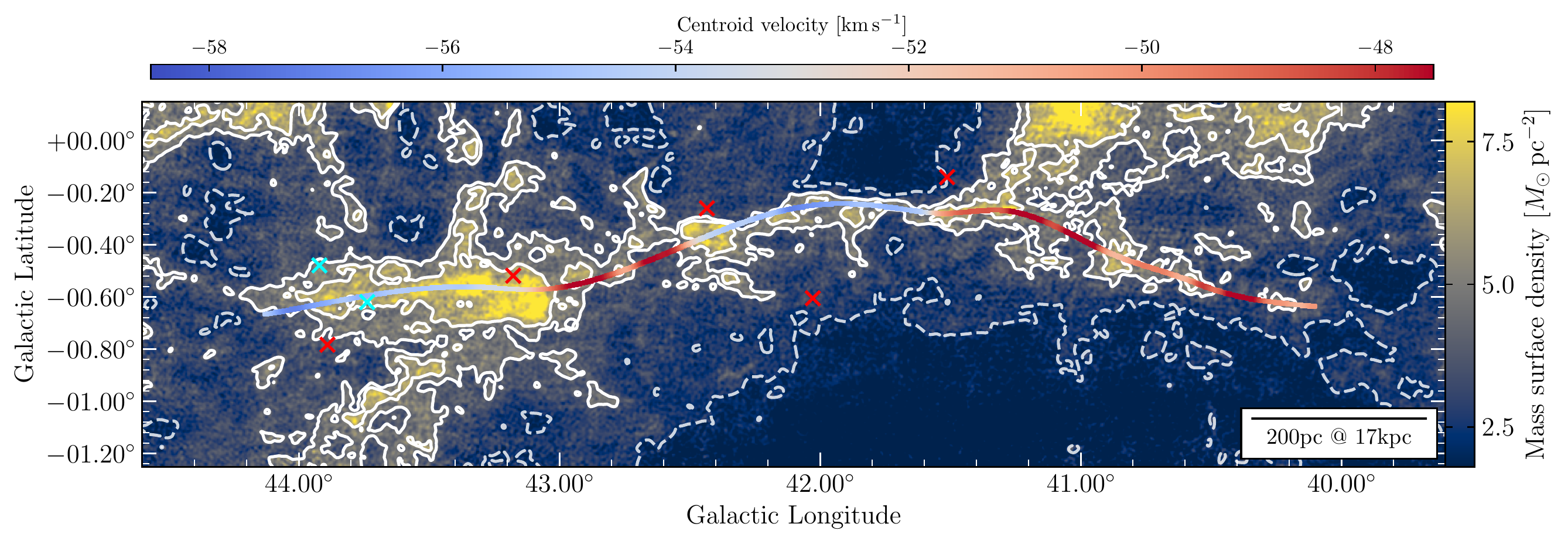}
      \caption[]{Mass surface density map. The map shows the \ion{H}{i} mass surface density integrated over the velocity interval $-57.5 < \varv_{\mathrm{LSR}} < -48.5\rm\,km\,s^{-1}$. The dashed contour gives the $5\sigma$ level at $2.3\,M_{\odot}\rm\,pc^{-2}$ by which the diffuse mass is calculated. The white contours show the mass surface densities at the levels $4.2$ and $5.6\,M_{\odot}\rm\,pc^{-2}$, respectively. The spine of Maggie is color-coded by the centroid velocity shown in Fig.~\ref{fig:pv_filament}. The crosses mark the positions of the continuum sources listed in Tab.~\ref{tab:continuum_sources}. The cyan crosses indicate the positions of absorption detections.}
      \label{fig:Maggie_massdens_map}
   \end{figure*}

    Figure~\ref{fig:Maggie_massdens_map} shows the mass surface density map of Maggie derived from the observed emission data. The mass surface density reaches values of more than $\sim$8.0$\,M_{\odot}\,\mathrm{pc}^{-2}\,(=1.0\times 10^{21}\,\mathrm{cm}^{-2},\,A_V\sim 1\rm\,mag)$\footnote{We relate the total hydrogen column density $N_\mathrm{H}=N_{\mathrm{HI}}+2\,N_{\mathrm{H_2}}$ to the visual extinction using $N_\mathrm{H}=2.2\times 10^{21}\,\mathrm{cm}^{-2}\,\mathrm{mag}^{-1}\times A_V$ \citep{2009MNRAS.400.2050G}. We take into account the molecular column density using the upper limit estimated in Sect.~\ref{sec:molecular_hydrogen_etc}.} toward the eastern hub and its distribution is constrained to within one order of magnitude (see Sect.~\ref{sec:coldens_PDF}). \citet{2003ApJ...587..278W} predict an average midplane thermal pressure at the Galactocentric radius $R_{\mathrm{GC}}=12\rm\,kpc$ of $P_{\mathrm{th,ave}}/k\sim 1580\rm\,K\,cm^{-3}$. We can estimate the expected length scale of Maggie along the line of sight by $l_{\mathrm{los}}\sim N_{\ion{H}{i}}\,T_{\mathrm{k}}/(P_{\mathrm{th,ave}}/k)$, where we set the kinetic temperature to be close to the density-weighted mean spin temperature $T_{\mathrm{k}}\approx \langle T_{\mathrm{s}}\rangle$. Since Maggie is significantly below the midplane, the pressure is likely to be smaller than $1580\rm\,K\,cm^{-3}$. Thus, we assume a canonical pressure of $P_{\mathrm{th,ave}}/k\sim 1000\rm\,K\,cm^{-3}$. At the mean column density of $\langle N_{\ion{H}{i}}\rangle=4.8\times 10^{20}\rm\,cm^{-2}$ (see also Sect.~\ref{sec:coldens_PDF}), the line-of-sight length is estimated to be $\sim$22$\rm\,pc$, which is roughly consistent with the on-sky width of 40\,pc given the uncertainty in temperature and pressure.
    
    Because of the diffuse nature of the WNM that is significantly contributing to the observed \ion{H}{i} emission it is difficult to define a clear edge of the filament. We therefore estimate two different masses: a ``diffuse mass'' in which the dense filament is embedded (mass above $5\sigma$), and a ``dense filament mass'' which resides within an approximately closed contour at the $9\sigma$ level ($=4.2\,M_{\odot}\rm\,pc^{-2}$) that roughly corresponds to the filament width obtained in Sect.~\ref{sec:location}.
    
    The total diffuse mass including the dense filament is $1.7\times 10^6\,M_{\odot}$. As mentioned above, due to its diffuse nature the mass derivation of the \ion{H}{i} has a high uncertainty as it is dependent on the selection of filament regions taken into account. Varying the contour level by $\pm 1\sigma$ to estimate the uncertainty, the dense filament mass is $7.2\substack{+2.5 \\ -1.9}\times 10^5\,M_{\odot}$. The atomic mass of Maggie therefore is comparable to the highest masses of large-scale filaments identified in the Milky Way \citep{2014A&A...568A..73R,2016A&A...590A.131A,2015MNRAS.450.4043W,2016ApJS..226....9W,2015ApJ...815...23Z,2018ApJ...864..153Z}, providing a large atomic gas reservoir to potentially host molecular cloud formation.
    
    If we take the column density we use to define the edge of the ``dense'' filament and assume that the width of the filament is the same along the line of sight as in the plane of the sky, we obtain a density of $4.2/40\,M_{\odot}\rm\,pc^{-3}\sim 7\times 10^{-24}\rm\,g\,cm^{-3}$, which equates to an H nucleus number density of roughly $4\rm\,cm^{-3}$. This is intermediate between the average densities we expect for the WNM and CNM at this Galactocentric radius \citep{2003ApJ...587..278W}, which is best explained by the filament being a mix of both components.
    
    However, given a mean density of $\bar n=4\rm\,cm^{-3}$, we point out that either the CNM mass fraction $f_{\mathrm{CNM}}$ must be higher or that a substantial fraction of the WNM must be in an unstable phase to be in agreement with the theoretical model given in \citet{2003ApJ...587..278W}. This follows from the mass conservation:
    \begin{equation}
        \bar n = f_{\mathrm{V,CNM}}\,n_{\mathrm{CNM}} + (1-f_{\mathrm{V,CNM}})\,\gamma\, n_{\mathrm{CNM}} \: ,
    \end{equation}
    \noindent where $f_{\mathrm{V,CNM}}$ is the volume filling fraction of the CNM, $n_{\mathrm{CNM}}$ is the CNM number density, and $\gamma=n_{\mathrm{WNM}}/n_{\mathrm{CNM}}=T_{\mathrm{CNM}}/T_{\mathrm{WNM}}\sim 0.01$ is the ratio of the CNM and WNM temperature in pressure equilibrium. The CNM mass fraction $f_{\mathrm{CNM}}=f_{\mathrm{V,CNM}}\,n_{\mathrm{CNM}}/\bar n$ can then be expressed as
    \begin{equation}
        f_{\mathrm{CNM}} = \frac{1}{1-\gamma}\,\left(1-\gamma\,\frac{n_{\mathrm{CNM}}}{\bar n}\right) \: .
    \end{equation}
    \noindent Assuming the \citet{2003ApJ...587..278W} CNM-WNM model, with $\bar n=4\rm\,cm^{-3}$ the CNM mass fraction is higher than our assumed values. For example, even for a CNM density $n_{\mathrm{CNM}}=40\rm\,cm^{-3}$ \citep[the maximum CNM density at $R_{\mathrm{GC}}=11\rm\,kpc$;][]{2003ApJ...587..278W}, the CNM fraction is still $f_{\mathrm{CNM}}=0.91$. The mean density can be brought into agreement with the previously derived CNM fraction of $\sim$0.5 if the temperature ratio becomes $\gamma\sim0.05$, which corresponds to a mean temperature in the unstable regime $\sim$1000--2000$\rm\,K$, compared to the higher temperatures of the classical WNM around $\sim$6000\,K. Observational works have found the fraction of \ion{H}{i} in the unstable phase to be at least 20--50\% \citep{2003ApJ...586.1067H,2013MNRAS.436.2366R,2015ApJ...804...89M,2018ApJS..238...14M,2019ApJ...880..141N}, possibly even higher \citep{2019MNRAS.483..593K}.
    
%-----------------------------------------------------------------

\section{Discussion}\label{sec:discussion}

\subsection{Molecular gas tracers, continuum, and dust}\label{sec:molecular_hydrogen_etc}

   What is the molecular gas fraction of Maggie and are there signatures of molecular cloud formation? To address this, we investigated the molecular lines \element[][12]{CO}, \element[][13]{CO}, and \element[18][]{C}O ($J=1$--0) using the Milky Way Imaging Scroll Painting \citep[MWISP;][]{2019ApJS..240....9S} survey. The rms noise of the data is $0.5\rm\,K$ for \element[][12]{CO} at the velocity resolution of $0.16\rm\,km\,s^{-1}$ and $0.3\rm\,K$ for \element[][13]{CO} and \element[18][]{C}O at $0.17\rm\,km\,s^{-1}$ resolution. The spatial resolution in all three tracers is $\sim$50\arcsec.
   
   In all three tracers, we detect no molecular gas emission in the velocity range between $-57.5$ and $-48.5\rm\,km\,s^{-1}$. We show in Appendix~\ref{sec:MWISP_app} a map of the integrated \element[][12]{CO} emission as an example. To estimate the molecular hydrogen content, we used the CO-$\rm H_2$ conversion factor $X=1.8\times 10^{20}\rm\,cm^{-2}\,K^{-1}\,km^{-1}\,s$ \citep{2001ApJ...547..792D}. We determined an upper limit at $5\sigma$ for the molecular surface density of $\Sigma_{\rm H_2}\sim 13\,M_{\odot}\rm\,pc^{-2}$. Since the sensitivity limit is close to the observed \ion{H}{i} column densities, we cannot entirely dismiss the presence of CO-bright $\rm H_2$ gas.
   
   Indeed, \citet{2020PASJ...72...43N} identify molecular gas clumps toward Maggie within small-scale \ion{H}{i} clouds that they infer from a dendrogram analysis of the VGPS \ion{H}{i} data \citep{2006AJ....132.1158S}. The average size of the \ion{H}{i} clouds is 5\,pc. They derive the molecular mass by summing the brightness temperature of the FUGIN \element[][12]{CO} data \citep{2017PASJ...69...78U} within each \ion{H}{i} cloud, including voxels with a brightness temperature lower than the noise. In doing that, diffuse emission can be recovered in the same manner as stacking emission spectra. The total $\rm H_2$ mass of all 353 identified cloudlets that are within the dense filament contour at $4.2\,M_{\odot}\rm\,pc^{-2}$ (area filling factor $\sim$0.21) is $\sim$6$\times 10^4\,M_{\odot}$, only a small fraction of the \ion{H}{i} mass (0.08), with the cloudlets' average surface density being $\sim$2.5$\,M_{\odot}\rm\,pc^{-2}$ that is consistent with our upper limit. The mean molecular mass fraction of the \ion{H}{i} clouds is $\sim$0.2 of the total gas mass \citep{2020PASJ...72...43N}. Naturally, the \ion{H}{i} ``leaves'' of the dendrogram analysis sample the highest-density parcels of the medium, hence a portion of the total molecular mass outside the \ion{H}{i} clouds might be missed.
   
   In addition, CO might not always be a good tracer of molecular hydrogen as ``CO-dark'' $\rm H_2$ could account for a significant fraction of the total $\rm H_2$ \citep{2008ApJ...679..481P,2009ApJ...692...91G,2016A&A...593A..42T}, particularly at low column densities \citep{2008ApJ...680..428G,2011A&A...536A..19P}. The CO-dark fraction has furthermore been observed to increase with Galactocentric distance \citep{2013A&A...554A.103P}. Simulations of molecular gas in Milky Way type galaxies indicate that the CO-dark fraction is indeed higher in regions of lower surface density \citep{2014MNRAS.441.1628S}. CO observations at visual extinctions $A_{\mathrm{V}}\leq 3.5$ could underestimate the molecular hydrogen content by a significant fraction \citep{2011MNRAS.412..337G}.
   
   As OH may be a better molecular gas tracer than CO in atomic-molecular transition regions at low column densities \citep[e.g.][]{2017ApJ...839....8T}, we inspected the strongest OH line at $1667\rm\,MHz$ in the THOR OH data \citep[see][]{2018A&A...618A.159R}. The THOR OH data consist of VLA C-array observations only, so large-scale emission cannot be recovered due to the missing flux on short $uv$ spacings. However, we can derive an OH abundance from OH absorption measurements against continuum sources. At the velocity range of Maggie, no absorption is found against any of the continuum sources listed in Tab.~\ref{tab:continuum_sources}. We note that depending on the peak flux of the continuum source and opacity of the OH line, the detection limit could be above the expected OH column density. The presence of molecular gas traced by OH can therefore not be ruled out. Follow-up OH observations at higher sensitivity will address this in a future study.
   
   We investigated the possibility of Maggie being a foreground structure by examining the extinction of stars up to a distance of 5 kpc using the early installment of the third \textit{Gaia} data release \citep{2021A&A...649A...1G}. The $K_{\mathrm{s}}$-band extinctions are obtained using the Rayleigh-Jeans Colour Excess \citep[RJCE,][]{2011ApJ...739...25M} method which combines near- and mid- infrared data from the Two Micron All-Sky Survey \citep[2MASS;][]{2006AJ....131.1163S} and the Wide-Field Infrared Survey Explorer \citep[WISE;][]{2010AJ....140.1868W}, respectively. The \textit{Gaia} Archive\footnote{https://gea.esac.esa.int/archive/.} provides the crossmatch of \textit{Gaia} sources with both the 2MASS and WISE catalogues.
   
   In the presence of an obscuring medium, for example an atomic cloud, high extinction in the densest parts would prevent the detection of stars in the \textit{Gaia} sample, yet extinction toward the outer portions of the cloud can still be detected in order to trace its silhouette. Figure~\ref{fig:gaia_map} presents the obtained $K_{\mathrm{s}}$-band extinctions toward Maggie for sources up to a distance of 5\,kpc from us. If Maggie were located within 5\,kpc from the Sun, higher extinction stars would represent its shape in the $\ell$-$b$ plane. However, since no signature of Maggie emerges from the extinction map, this indicates that Maggie is not a foreground cloud with large deviation from the Galactic rotation velocity.
   
   We furthermore detect neither extinction features in the Spitzer Galactic Legacy Infrared Mid-Plane Survey Extraordinaire  \citep[GLIMPSE;][]{2009PASP..121..213C} that might hint at the presence of IRDCs nor any excess infrared emission or signatures of stellar activity as observed in all bands of GLIMPSE and the Improved Reprocessing of the IRAS Survey \citep[IRIS;][]{2005ApJS..157..302M}. There are two possible reasons for not detecting IRDC features: 1) Maggie has not formed cold, dense molecular regions on a large scale such that there exist no IRDCs within Maggie or 2) Maggie is too distant to be observed as an IRDC due to the lack of emission background.
   
   We have also examined the high-sensitivity maps of \textit{Planck} at 353, 545, and 857\,GHz emission \citep{2014A&A...571A...1P} and the higher angular resolution \textit{Herschel} Hi-GAL data \citep{2017MNRAS.471.2730M} to further search for signatures of Maggie. We do not detect Maggie in any of the continuum bands while noting that we are most likely dominated by foreground emission in the Galactic plane.
   
   After taking a variety of surveys into consideration, Maggie appears to show no signs of stellar activity as it is mostly atomic, while molecular gas has formed only in high-density clumps on the smallest spatial scales \citep{2020PASJ...72...43N}.
    
\subsection{Kinematic signatures}
\subsubsection{Mach number distribution}
   In the following, we adopt a stable two-phase CNM-WNM model to derive the turbulent Mach number distributions. We estimate the three-dimensional scale-dependent Mach number of the filament assuming isotropic turbulence with $\mathcal{M}=\sqrt{3}\,\sigma_{\mathrm{turb}}/c_s$, where $\sigma_{\mathrm{turb}}$ and $c_s$ are the turbulent one-dimensional velocity dispersion and sound speed, respectively.
   The turbulent line width is calculated by subtracting the thermal line width contribution from the observed line width as
   \begin{equation}
       \sigma_{\rm turb} = \sqrt{\sigma_{\rm obs}^2 - \sigma_{\rm th}^2} \: ,
   \end{equation}
   \noindent where $\sigma_{\rm obs}$, and $\sigma_{\rm th}$ are the observed, and thermal velocity dispersion, respectively.
   Since we expect the observed line width to be a combination of both CNM and WNM, and do not know the turbulent contribution to the observed line width of either \ion{H}{i} phase, we compute a single Mach number for the two-phase medium using a weighted mean of the thermal line width and sound speed. The thermal velocity dispersion can be decomposed into a CNM and WNM component as
   
   \begin{equation}
       \sigma_{\mathrm{th}}^2 = f_{\mathrm{CNM}}\,\sigma_{\mathrm{th,CNM}}^2 + (1-f_{\mathrm{CNM}})\,\sigma_{\mathrm{th,WNM}}^2 \: ,
   \end{equation}
   
   \noindent where $\sigma_{\mathrm{th,CNM}}$ and $\sigma_{\mathrm{th,WNM}}$ refer to the thermal velocity dispersion of the CNM and WNM, respectively. The combined velocity dispersion and sound speed can then be computed using the weighted mean kinetic temperature
   \begin{equation}
       \langle T_{\mathrm{k}}\rangle = f_{\mathrm{CNM}}\,T_{\mathrm{k,CNM}} + (1-f_{\mathrm{CNM}})\,T_{\mathrm{k,WNM}} \: ,
   \end{equation}
   
   \noindent where $T_{\mathrm{k,CNM}}$ and $T_{\mathrm{k,WNM}}$ are the kinetic temperatures of the CNM and WNM, respectively.
   
   We assume that the kinetic temperature of the CNM is close to the spin temperature and set $T_{\mathrm{s,CNM}}=T_{\mathrm{k,CNM}}$. In pressure equilibrium, \citet{2003ApJ...587..278W} predict a range of WNM kinetic temperatures between 4100--$8800\rm\,K$. As the densities of the WNM are low, the gas can usually not be thermalized by collisional excitation only, such that $T_{\mathrm{s,WNM}}<T_{\mathrm{k,WNM}}$. Through excitation by resonant Ly$\alpha$ photon scattering, sufficient Ly$\alpha$ radiation can allow $T_{\mathrm{s,WNM}}\to T_{\mathrm{k,WNM}}$. The WNM spin temperature $T_{\mathrm{s}}$ has been observed to be as high as $7200\rm\,K$ \citep{2014ApJ...781L..41M}. We therefore assume a mean WNM kinetic temperature of $T_{\mathrm{k,WNM}}=6000\rm\,K$.
   
   Since the thermal line width and sound speed scale as $T_{\mathrm{k}}^{1/2}$, the variation is moderate and does not change the Mach number significantly. We calculate three different Mach number distributions for $f_{\rm CNM}=30,\,50,\,70\%$, corresponding to $T_{\mathrm{k,CNM}}=40,\,70,\,100\rm\,K$.

\begin{figure}
      \centering
        \resizebox{\hsize}{!}{\includegraphics{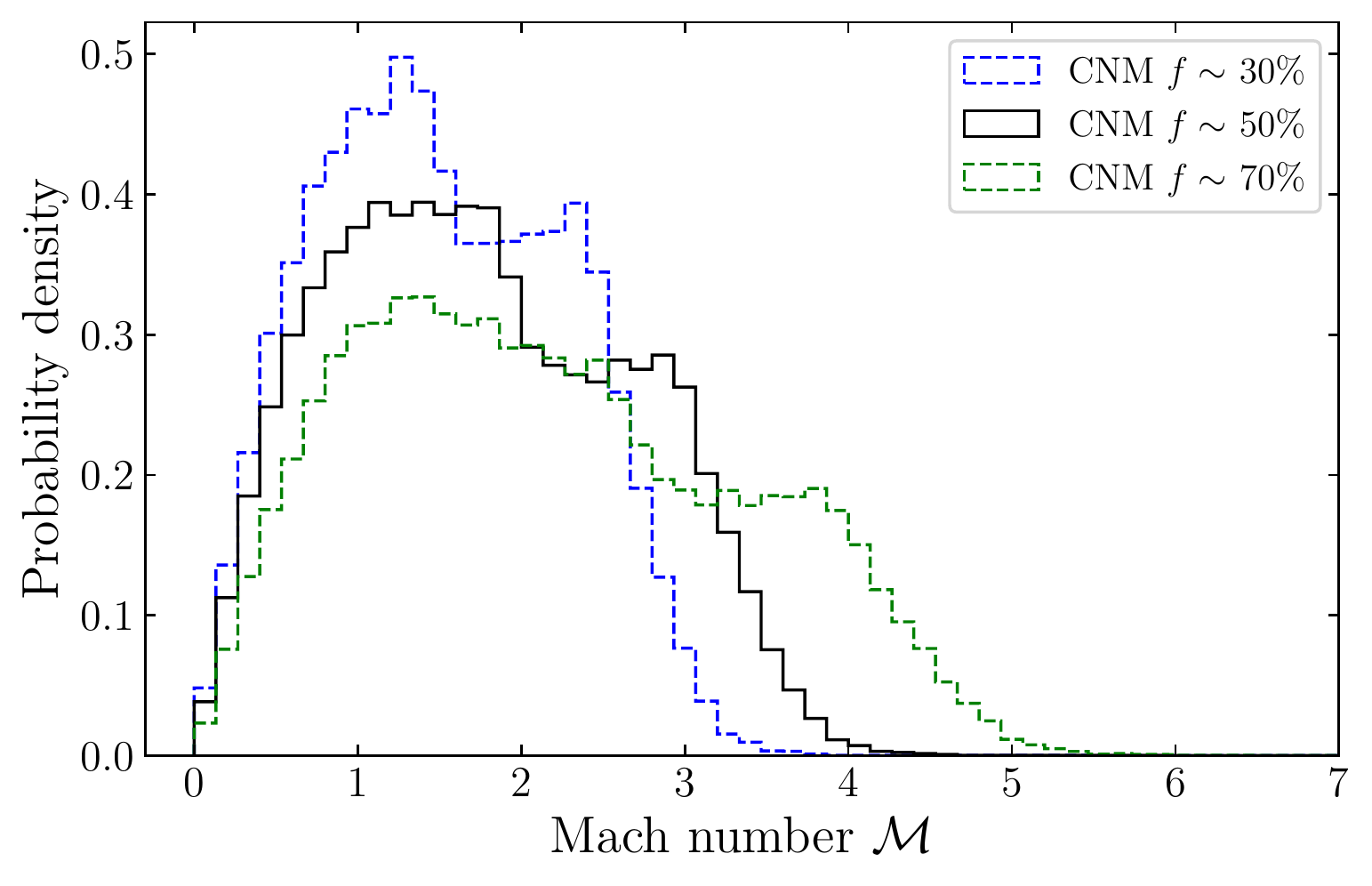}}
      \caption[]{Mach number distribution of Maggie. The blue, black, and green histograms show the Mach number distributions assuming a CNM fraction of 30, 50, and 70\%, respectively. Each distribution is computed using a single thermal line width and sound speed inferred from a weighted mean kinetic temperature.}
      \label{fig:mach_number}
    \end{figure}

   Figure~\ref{fig:mach_number} shows the Mach number distributions of Maggie under different assumptions of the CNM fraction. The median Mach number is 1.5, 1.7, and 2.0 for the CNM fractions 30, 50, and 70\%, respectively, which is generally in contrast to molecular filaments being driven by highly supersonic turbulence \citep[e.g.][]{2004ARA&A..42..211E}. However, since we use a single mean kinetic temperature in each case, the Mach number distributions also reflect a slight bimodality that is already evident in the line width distribution (Fig.~\ref{fig:hist_fwhm}). While each Mach number distribution represents a different mixture of CNM-WNM gas, we note that the bump toward higher Mach numbers reflected in each distribution is due to the increased line widths associated with higher $\varv_{\mathrm{LSR}}$ velocities, likely contaminated by blending background gas (see Sect.~\ref{sec:kinematics}).
   
   Numerical studies show that the condensation process toward a stable CNM phase depends on the properties of turbulence in the WNM \citep{2011A&A...526A..14S,2014A&A...567A..16S,2020A&A...643A..36B}. These studies suggest that highly supersonic turbulence that is driven by compressive modes regulates the formation of a stable CNM phase as turbulent compression and decompression of the gas will be much more rapid than the cooling timescale. Once the gas has moved toward an unstable phase, that is the cooling timescale $t_{\rm cool}$ is shorter than the dynamical time $t_{\rm dyn}=\ell / c_s$, the gas can cool down fast enough to reach a stable branch of the CNM before being decompressed again \citep{1999A&A...351..309H}.
   On the other hand, a too low degree of turbulence cannot produce sufficient density fluctuations to sustain the formation of a thermally stable CNM. The WNM is mostly found in a sub-/transonic regime \citep[e.g.][]{2020arXiv201203160M}, therefore providing the dynamical conditions for thermal instability to grow efficiently to form a bistable \ion{H}{i} phase.
   
   Absorption studies show that a substantial fraction of \ion{H}{i} in fact lies in a thermally unstable or transient phase \citep{2001ApJ...551L.105H,2003ApJ...586.1067H,2013MNRAS.436.2352R,2013MNRAS.436.2366R}. Shifting the mean temperature of the WNM to an unstable regime around $\sim$1500\,K leads to Mach numbers in the range $2-4$.

\subsubsection{Structure function}\label{sec:structure_function}
\begin{figure}
      \centering
        \resizebox{\hsize}{!}{\includegraphics{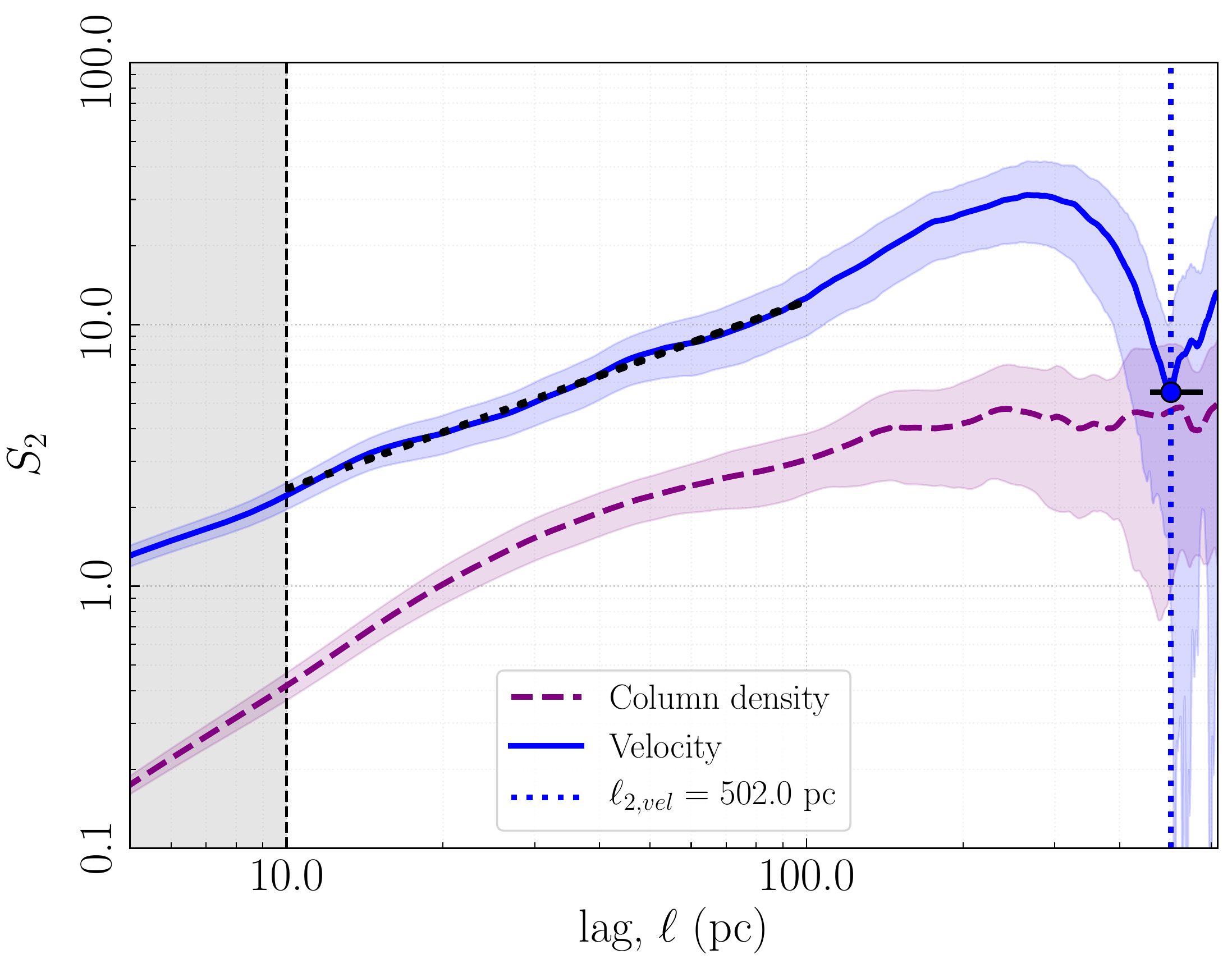}}
      \caption[]{Second order structure function. This plot shows the second order structure function of column density (dashed purple) and velocity (blue) computed from the respective average quantity along the Maggie filament. A 2-D plane fit to the velocity field has been subtracted prior to computing the velocity structure function. The vertical blue dotted line indicates the characteristic spatial scale of the velocity. The black dotted line is a power-law fit to the velocity structure function. The gray shaded area indicates the range of scales below three beams for which the recovered structure function is unreliable.}
      \label{fig:structure_function}
    \end{figure}

   In order to investigate the velocity and column density structure of Maggie in a statistical sense, we compute the 1-D structure function \citep[see][]{2020NatAs...4.1064H} of the data using
   \begin{equation}
       S_p(\ell) = \langle \delta x(r,\ell)^p \rangle = \langle \vert x(r) - x(r+\ell) \vert^p \rangle \: ,
   \end{equation}
   \noindent where $\delta x(r,\ell)$ is the absolute difference in the quantity $x$ measured between two locations $r$ and $r+\ell$ separated by $\ell$. The brackets indicate the average of the quantity over all locations. The order $p$ of the structure function can be directly related to a physical quantity. We use the second order structure function, which in the case of the velocity structure is proportional to the kinetic energy of the line-of-sight component of the flow.
   We show in Fig.~\ref{fig:structure_function} the second order structure function derived from the weighted centroid velocity (Fig.~\ref{fig:pv_filament}), after having subtracted a 2-D plane fit to the velocity field in order to better examine the wave-like oscillation, and the column density along the $40\rm\,pc$-wide spine of Maggie. A significant local minimum is evident at the spatial scale of $\ell_{2,vel}\approx 500\rm\,pc$. Since the structure function is evaluated for pairs of positions that are separated by the lag $\ell$, the maximum spatial scale at which the structure function can be fully sampled is given by half the total length of the filament. Therefore, the structure function is only computed over spatial scales $\ell \lesssim 600\rm\,pc$. Consequently, the full velocity `dip' cannot be recovered and we estimate the uncertainty of the scale by the difference between $\ell_{\mathrm{max}}$ and $\ell_{2,vel}$, giving $\ell_{2,vel}=502 \pm 113\rm\,pc$. This result confirms the wave-like pattern seen in Figs.~\ref{fig:Maggie_overview} and \ref{fig:pv_filament}.
   
   The column density structure of Maggie does not show any characteristic length scale at which the structure function has a local minimum but roughly represents a power-law scaling, indicative of the scale-invariant behavior of a gas structure generated by turbulence \citep{2004ARA&A..42..211E}.
   
   The phenomenological Kolmogorov relation for incompressible turbulence predicts a dissipationless cascade of energy in the inertial subrange, such that the kinetic energy transfer rate $\mathcal{E}\approx \delta \varv^3/\ell$ stays constant \citep{1941DoSSR..30..301K}. Due to the nature of the turbulent cascade, the velocity dispersion then increases with scale as $\delta \varv \propto \ell^{q_{\mathrm{theo}}}$, where $q_{\mathrm{theo}}=1/3$, which we can relate to the power-law scaling of the second order structure function $\langle \delta \varv^2 \rangle \propto \ell^\gamma$, where $\gamma = 2q_{\mathrm{theo}}$ \citep[see e.g.][]{2004ARA&A..42..211E,2020NatAs...4.1064H}.
   We therefore fitted the velocity structure function with a power-law function of the form $S_2(\ell) \propto \ell^\gamma$. We set the lower limit of the fitting range to three beam sizes (10\,pc), below which no reliable structure functions can be recovered, and the upper limit at the approximate scale above which the structure function begins to turn over (100\,pc). For Maggie we find $q=0.36\pm 0.02$ in the case where we subtracted the 2-D plane fit, which is in good agreement with the expected power-law scaling $q_{\rm theo}=1/3$ in the sub-/transonic regime \citep[see also][]{2003ApJ...587..278W,2004ARA&A..42..211E,2007ARA&A..45..565M,2014A&A...567A..16S,2020arXiv201203160M}. While the dip in the velocity structure function is less pronounced if we do not subtract a 2-D plane fit, the power-law scaling remains the same with $q=0.37\pm 0.02$.
   These values are in agreement with the observational scaling exponents originally found by \citet{1979MNRAS.186..479L,1981MNRAS.194..809L}, who investigated the velocity dispersion--size relation in a sample of molecular clouds and clumps using different tracers. He utilized the radial velocity dispersion mainly in terms of the line width of optically thin lines, such that the relation between line width and individual cloud size can be interpreted as a correlation between velocity fluctuation and scale size \citep[see also][]{1984ApJ...277..556S}. \citet{1979MNRAS.186..479L,1981MNRAS.194..809L} found a velocity dispersion--size relation following a power law with an index $q_{\mathrm{L}}=0.38$, which is, although slightly larger, similar to the Kolmogorov relation. Systematic studies of more homogeneous samples of molecular clouds traced in CO found velocity scaling exponents in a range between 0.24 and 0.79 (see Table 2 in \citet{2021MNRAS.500.5268I}, which summarizes more than 40 years of observational work).
   
   \citet{2021MNRAS.500.5268I} find in their simulations that the scaling exponents lean toward lower values when molecular clouds are dominated by the influence of the Galactic potential rather than the effects of clustered SN feedback and self-gravity \citep[see also][]{2019A&A...630A..97C}. This suggests that low velocity scaling exponents are associated with diffuse regions of low star forming activity \citep[e.g.][]{2010A&A...512A..81F}, which also favors the development of long filamentary structures \citep{2020MNRAS.492.1594S}.
   
   Maggie's velocity scaling is in better agreement with the lower end of the values found for molecular clouds (see above). This regime around 1/3 is indicative of Kolmogorov turbulence, and following \citet{2021MNRAS.500.5268I}, this turbulence may be caused by the Galactic potential rather than being driven by the effects of self-gravity or stellar feedback.
   
   The existence of density enhancements that are spaced at a characteristic scale within a highly filamentary cloud would be in broad agreement with theoretical predictions of the fragmentation of a self-gravitating fluid cylinder due to the ``sausage'' instability \citep[e.g.][]{1953ApJ...118..116C}. As we find no characteristic scale in the column density structure, we argue that the effects of self-gravity have not yet become important and it is difficult to correlate the observed velocity signature with possible mechanisms driving the column density structure formation.  
   
   We caution, as pointed out in Sect.~\ref{sec:coldens_mass}, that the column density is a product of both the CNM and WNM and is inferred from the integration over velocity. This could lead to a structure that smooths over an underlying higher-density CNM structure due to the diffuse nature of the WNM. The column density structure can furthermore not be described by single power-law alone. This could be due to the filament being a mix of both CNM and WNM and that the two phases have different turbulence scalings. With the WNM being more widespread, the structure function as probed by \ion{H}{i} emission could be dominated by the warm diffuse gas on large scales while the CNM dominates the power-law on small scales \citep[see e.g.][]{2019MNRAS.483.3437C}. The integrated emission might therefore not be the best estimator of the spatial distribution of the column density.
   
\subsection{Column density PDF}\label{sec:coldens_PDF}
   What are the dominant physical processes acting within the cloud? To elaborate on the density structure of Maggie, we employ the commonly used probability density function (PDF) of the column density. The shape of column or volume density PDFs are used as a means to describe the underlying physical mechanisms of a cloud \citep[e.g.,][]{2013ApJ...763...51F,2014Sci...344..183K,2015A&A...575A..79S}. A log-normal shape of a PDF is considered a signature of turbulent motion dominating a cloud's structure \citep[see e.g.][]{2000ApJ...535..869K}. The magnitude of turbulence can furthermore be reflected in the width of a log-normal PDF and can then be associated with the Mach number \citep[e.g.,][]{1997MNRAS.288..145P,1998PhRvE..58.4501P,2007ApJ...665..416K,2008PhST..132a4025F,2012ApJ...761..149K,2012MNRAS.423.2680M}, while noting that the turbulence driving scale and CNM-WNM mass ratio also affect the width of the PDF \citep{2017ApJ...843...92B}.
   
   Molecular clouds that are subject to the increasing effect of self-gravity develop high-density regions, producing a power-law tail in their PDF \citep[e.g.][]{2014ApJ...781...91G,2017ApJ...834L...1B}. Many star-forming molecular clouds have been confirmed to show such power-law tails \citep{2009A&A...508L..35K,2013ApJ...766L..17S,2016A&A...587A..74S}. Even before the effects of gravity become dominant, gravitationally unbound clumps can exhibit power-law tails due to pressure confinement from the surrounding medium \citep{2011A&A...530A..64K}.

\begin{figure*}
      \centering
        \includegraphics[width=1.0\textwidth]{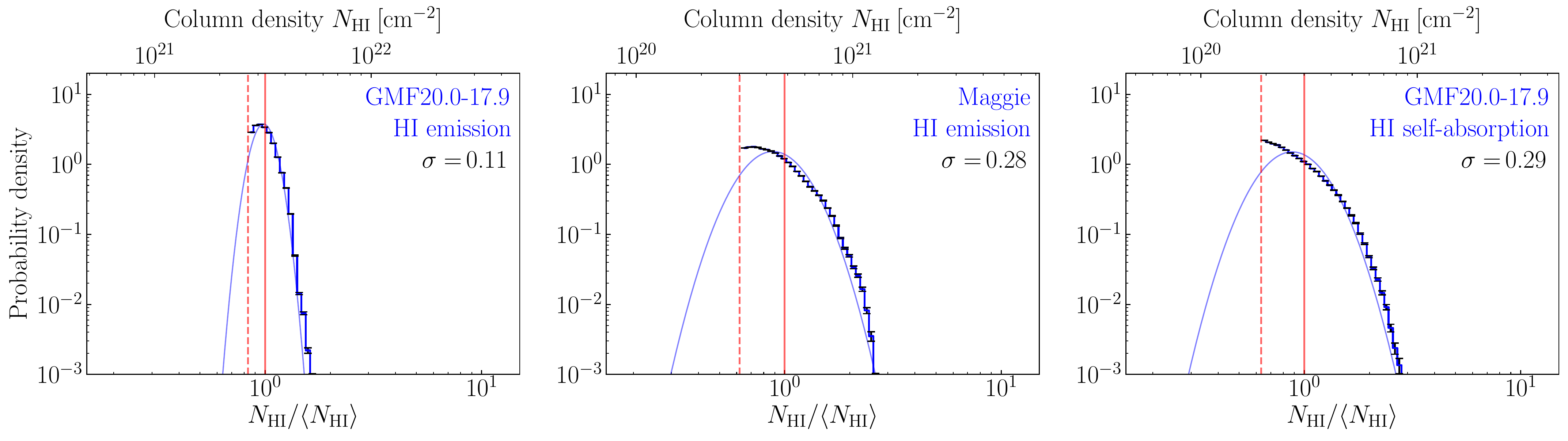}
      \caption[]{Comparison of N-PDFs. The blue histograms show the column density distributions of Maggie (\textit{middle}) in comparison with the N-PDFs taken from a different GMF region analyzed in \citet{2020A&A...642A..68S} that trace the midplane \ion{H}{i} emission (\textit{left}) and CNM by means of \ion{H}{i} self-absorption (\textit{right}). The blue curve in each panel indicates a log-normal fit to the distribution. The red vertical dashed and solid lines mark the column density thresholds taken into account and the mean column density, respectively.}
      \label{fig:N-PDF}
    \end{figure*}

   We show in Fig.~\ref{fig:N-PDF} the normalized column density PDF (N-PDF) of Maggie taken above the $5\sigma$ limit at $\sim$3$\times 10^{20}\rm\,cm^{-2}$ ($=2.3\,M_{\odot}\rm\,pc^{-2}$). The mean of the N-PDF is $4.8\times 10^{20}\rm\,cm^{-2}$. The distribution varies within one order of magnitude and can be well described by a log-normal function of width $\sigma=0.28$. We caution that the shape and width of the resulting N-PDFs are sensitive to the region taken into account \citep{2015A&A...575A..79S,2015A&A...576L...1L,2016A&A...590A.104O,2019MNRAS.482.5233K}. If we only consider column densities within an approximately last closed contour at $\sim$4$\times 10^{20}\rm\,cm^{-2}$, the N-PDF shows a narrower log-normal shape with a width smaller by $\Delta\sigma=0.05$. 
   
   Figure~\ref{fig:N-PDF} presents a comparison between the N-PDF of Maggie and the optical depth corrected N-PDFs derived from \ion{H}{i} emission and self-absorption toward the giant molecular filament GMF20.0-17.9 \citep{2020A&A...642A..68S}. The width of Maggie's N-PDF comes close to the value found for the CNM gas traced by HISA and is larger than that of the atomic gas traced by \ion{H}{i} emission toward the midplane \citep[see also][]{2020A&A...634A.139W}. However, the width of the CNM distribution is likely to be higher as the HISA N-PDF is limited by the column density range at which \ion{H}{i} self-absorption can be detected. 
   
   Since we use the same data set and follow the same methodology for deriving the N-PDF as \citet{2020A&A...634A.139W} and \citet{2020A&A...642A..68S}, systematic differences should be minimized. Despite that, we note a slight cutoff in the PDF at $N_{\ion{H}{i}}\sim 1\times 10^{21}\rm\,cm^{-2}$. As we use a uniform optical depth correction for the whole cloud (Sect.~\ref{sec:coldens_mass}), the optical depth could be underestimated in regions of high column density. If we assume a column density $\sim$10$^{21}\rm\,cm^{-2}$ and a FWHM line width $\sim$10$\rm\,km\,s^{-1}$, the peak optical depth becomes $\tau\sim 1$. The cutoff in the PDF could therefore be due to the cloud becoming optically thick and the width could then be underestimated.
   
   Narrow log-normal shaped N-PDFs are commonly observed in the diffuse \ion{H}{i} emission toward well-known molecular clouds \citep{2015ApJ...811L..28B,2016ApJ...829..102I,2017MNRAS.472.1685R}. On the other hand, the HISA N-PDFs that trace the CNM show similar or somewhat broader distributions, indicative of the clumpy structure and higher degree of turbulence. The CNM-WNM mixture of Maggie, that is spatially more clearly defined than the material usually traced by \ion{H}{i} emission, agrees with our previous findings and suggests that Maggie might be at the verge of becoming a dense filament forming out of the diffuse atomic ISM.
    
\subsection{Molecular gas formation and timescale}\label{sec:atomictransition}
   Is Maggie expected to form molecular gas at its current evolutionary stage? We evaluate the surface density threshold of \ion{H}{i} required to efficiently form molecular hydrogen, based on the analytical models by \citet{2014ApJ...790...10S}, and \citet{2016ApJ...822...83B}. These authors derive a steady-state \ion{H}{i}-to-$\rm H_2$ transition model assuming a slab irradiated isotropically on two sides by FUV flux. The expected \ion{H}{i} surface density at the transition then is
   \begin{equation}
       \Sigma_{\ion{H}{i}} = \frac{6.71}{\tilde{\sigma}_g} \, \mathrm{ln}\,\Bigg(\frac{\alpha G}{3.2} + 1 \Bigg) \: M_{\odot}\,\mathrm{pc^{-2}}
       \label{equ:Sigma_HI}
   \end{equation}
   \noindent \citep[see Eq.~(1) in][]{2017ApJ...835..126B}, where $\tilde{\sigma}_g$ is the dust absorption cross section per hydrogen nucleus in the Lyman-Werner (LW) dissociation band (11.2--13.6$\rm\,eV$) relative to the standard solar neighborhood value, $\alpha$ is the ratio of the unshielded $\rm H_2$ dissociation rate to formation rate, and $G$ is the average $\rm H_2$ self-shielding factor in dusty clouds. The product $\alpha G$ is the $\rm H_2$ formation-to-destruction rate ratio, accounting for $\rm H_2$ shielding by $\rm H_2$ line absorption and dust, and may be written as
   \begin{equation}
       \alpha G = 2.0 \, I_{\mathrm{UV}} \, \Bigg( \frac{30\,\mathrm{cm^{-3}}}{n} \Bigg) \, \Bigg( \frac{9.9}{1 + 8.9\,\tilde{\sigma}_g} \Bigg)^{0.37}
       \label{equ:aG}
   \end{equation}
   \noindent \citep[][]{2014ApJ...790...10S,2016ApJ...822...83B}, where $I_{\mathrm{UV}}$ is the free-space (unshielded) interstellar radiation intensity relative to the \citet{1978ApJS...36..595D} field, and $n$ is the number density. The UV intensity exponentially decreases with Galactocentric distance \citep{2003ApJ...587..278W}. For $R_{\mathrm{GC}}=12\rm\,kpc$, \citet{2003ApJ...587..278W} suggest $I_{\mathrm{UV}}\sim 0.39$. We note that this approximation is derived for the midplane and should be roughly constant up to the scale height of OB stars and could be significantly lower at larger distances from the plane. Since Maggie's distance from the plane is an order of magnitude greater than the OB scale height in the solar neighborhood \citep{2000AJ....120..314R}, the UV intensity should be viewed as an upper limit with a relative uncertainty of at least 40\% \citep[see also][]{2006AJ....131.2700E}
   
   The dust absorption cross-section $\tilde{\sigma}_g$ can be estimated as the inverse of the gas-to-dust ratio at the position of Maggie relative to the value in the local ISM if we assume that the dust grain properties are the same in the outer Galaxy. \citet{2017A&A...606L..12G} derive a gas-to-dust ratio $\gamma$ as a function of galactocentric distance and metallicity based on C\element[][18]{O} observations. Their reported gas-to-dust gradient predicts $\gamma\sim305$ for Maggie and $\sim$141 as a local value for $R_{\mathrm{GC}}=8.15\rm\,kpc$ \citep{2019ApJ...885..131R}. The relative dust absorption cross-section then becomes $\tilde{\sigma}_g=0.46$. With the estimated number density of $\bar n=4\rm\,cm^{-3}$ (see Sect.~\ref{sec:coldens_mass}), the \ion{H}{i} threshold for molecular hydrogen to form efficiently is expected at $\Sigma_{\ion{H}{i}}\approx 17.5\,M_{\odot}\rm\,pc^{-2}$. This exceeds the observed column densities by a factor of more than 2.
   
   However, we note that the model by \citet{2014ApJ...790...10S} is highly idealised and the derived gas-to-dust relation in \citet{2017A&A...606L..12G} has large systematic uncertainties due to variations in the CO abundance and poorly constrained dust properties. In addition with the uncertainties in distance and UV intensity, the \ion{H}{i} threshold is uncertain by at least $\sim$50\%.
   
   Furthermore, observational effects might produce a bias against the measurement of a clumpy multiphase medium. Small pockets of high-density CNM clumps could be hidden below the resolution of the telescope beam ($\approx$3.3\,pc). Even if we assume that those unresolved CNM clumps have a small volume filling fraction $f_{\mathrm{V,CNM}}$, a substantial fraction of the mass $f_{\mathrm{CNM}}=(n_{\mathrm{CNM}}\cdot f_{\mathrm{V,CNM}})/\bar n$ could be allocated in the CNM phase although the observed mean density is low compared to that of the CNM. High-density CNM clumps are much more efficient at shielding and allow the onset of $\rm H_2$ formation at lower surface densities. As we argue in Sect.~\ref{sec:coldens_mass}, the CNM fraction should be considerable and could allow $\rm H_2$ formation on a scale that is inaccessible to our observations.
   
   In fact, \citet{2020PASJ...72...43N} report a total molecular gas mass of $\sim$6$\times 10^4\,M_{\odot}$ that resides within small CNM cloudlets with an average size of 5\,pc (see Sect.~\ref{sec:molecular_hydrogen_etc}). These cloudlets have a mean \ion{H}{i} density of $\sim$50\,$\rm cm^{-3}$. Inserting this density into Eqs.~\eqref{equ:aG} and \eqref{equ:Sigma_HI} gives a molecule formation threshold at $\Sigma_{\ion{H}{i}}\approx 2.5\,M_{\odot}\rm\,pc^{-2}$, which is below their observed \ion{H}{i} surface density of $\sim$7\,$M_{\odot}\rm\,pc^{-2}$. Hence, we would expect molecule formation to take place within these high-density CNM cloudlets. Clearly, employing segmentation methods such as the dendrogram algorithm rather than using integrated quantities is important when aiming at identifying small-scale CNM structures in a clumpy multiphase medium.
   
   We estimate the timescale of $\rm H_2$ formation by the time after which the balance between $\rm H_2$ destruction and formation rate has reached a steady state. If we consider the dissociation rate to be negligible compared to the formation rate, the $\rm H_2$ formation timescale is given by
   \begin{equation}
       t_{\mathrm{H_2}}\sim \frac{1}{2\,R\,n}
   \end{equation}
   \noindent \citep[see Eq.~(23) in][]{2016ApJ...822...83B}, where $R=3\times 10^{-17}\,\tilde{\sigma}_g\rm\,cm^{3}\,s^{-1}$ is the $\rm H_2$ formation rate coefficient and $n$ is the total hydrogen number density that is essentially given by our estimated \ion{H}{i} number density in the absence of $\rm H_2$. For a mean density of $\bar n=4\rm\,cm^{-3}$ the formation timescale is $\sim$300\,Myr, about an order of magnitude larger than those predicted by models of nearby molecular clouds \citep{2007ApJ...654..273G} and numerical simulations \citep{2012MNRAS.424.2599C}. As the formation of molecular hydrogen is most effective on the surface of dust grains \citep{1963ApJ...138..393G}, the formation timescale at a given density is primarily limited by the dust-to-gas ratio $\propto\tilde{\sigma}_g$, rendering the formation of molecular hydrogen less likely in the outer Galaxy. However, the effects of small-scale density fluctuations, either due to a clumpy CNM-WNM medium or due to transient structures produced by supersonic turbulence \citep{2007ApJ...659.1317G}, can shorten the timescale. For CNM clumps at several times the observed density, the timescale could decrease significantly. Again using the \ion{H}{i} density reported by \citet{2020PASJ...72...43N}, the molecule formation timescale within the CNM-rich structures becomes $\sim$25\,Myr.
   
   In conclusion it is difficult to tightly constrain the formation of molecules within Maggie based on integrated \ion{H}{i} and CO observations alone. Because of observational effects small-scale density fluctuations might be missed that allow the formation of molecular hydrogen even at the current stage. On the smallest scales within CNM-rich structures, the presence of molecular hydrogen traced by CO emission has been confirmed by \citet{2020PASJ...72...43N}. In addition, time delays between the formation of $\rm H_2$ and \element[][]{CO} could lead to Maggie not being identified as a molecular cloud, even though the formation of molecular hydrogen is already taking place on larger scale \citep{2012MNRAS.424.2599C}.

\subsection{Comparison with literature}

   Maggie's origin remains unclear. While there are morphological and kinematic similarities, it is difficult to argue in favor of Maggie being an immediate counterpart to large-scale molecular filaments known in the literature.
   \begin{itemize}
       \item The aspect ratio of Maggie is in rough agreement with large-scale filaments and Milky Way bones \citep{2015MNRAS.450.4043W,2015ApJ...815...23Z}, even though we point out that the determination of filament widths is in many cases subject to the tracer being used to identify the filament.
       \item Maggie has a smooth velocity gradient of less than $\pm 3\rm\,km\,s^{-1}\,(10\,pc)^{-1}$ (Sect.~\ref{sec:kinematics}) and therefore satisfies the Milky Way bone criteria applied in \citet{2015ApJ...815...23Z}.
       \item Maggie does not display any close association with a spiral arm structure although the predicted location of the spiral arms is highly model-dependent \citep[see e.g.][]{2015ApJ...815...23Z}. It is observed that all types of large-scale filaments identified in the literature commonly show a range of orientations with respect to spiral arm features in position-velocity ($\ell$-$v$) space, both tracing out the spine of spiral arms as well as following $\ell$-$v$ tracks that are inclined to spiral arms \citep{2018ApJ...864..153Z}. The latter could potentially trace a spurious feature trailing off a spiral arm.
       \item The orientation of Maggie's major axis is remarkably well aligned with the Galactic midplane, a fact that the majority of large-scale filaments have in common \citep{2018ApJ...864..153Z}. However, Maggie is observed at a large distance of $\approx$500$\rm\,pc$ from the Galactic midplane while most filaments are found in spatial proximity to it. The fact that Maggie shows both a spatial (position-position) \textit{and} kinematic (position-velocity) displacement from any spiral structure suggests a different formation path than those of known molecular filaments in the disk.
       \item The Radcliffe wave \citep{2020Natur.578..237A} that consists of an association of molecular cloud complexes in the solar neighborhood has an aspect ratio of 20:1 and exhibits an undulating structure (in $p$-$p$-$p$ space) below and above the midplane. At its peak, it reaches an offset of $\sim$200\,pc from the plane. It is argued that the accretion of a tidally stretched gas cloud settling into the Galactic disk could in principle mimic the shape of the observed structure, but seems unlikely given that it would require the Radcliffe wave to be synchronized with the Galactic rotation over large scales.
   \end{itemize}
   Infalling gas, possibly of extragalactic origin, would interact with the gas in the disk and become compressed or even shocked along the edge closest to the disk if the relative velocities are large enough. In the case of Maggie, we then consider the chances of infalling gas becoming plane-parallel to be small. Additionally, in this scenario it seems unlikely for the gas to keep a coherent velocity structure.
   
   Conversely, gas that is being expelled out of the plane as a result of SN feedback tends to have a randomized orientation with respect to the Galactic disk \citep{2020MNRAS.492.1594S}. Observations of the preferentially vertical \ion{H}{i} filaments seen in the Galactic plane can be associated with enhanced SN feedback \citep{2020A&A...642A.163S} and indicate that Maggie is unlikely to have originated in SN feedback events.
   
   As Maggie could be linked to the global spiral arm structure in $\ell$-$\varv$ space, we speculate that the filament originated in the midplane and might have been brought to large distance by vertical oscillations. \citet{2006ApJ...643..881L} decomposed the structure of the Galactic \ion{H}{i} disk into Fourier modes. They found that the Galactic warp \citep[e.g.][]{1988gera.book..295B} can be described sufficiently well by the first three Fourier modes. Locally, significant high-frequency modes at large Galactocentric distances are also evident within the disk. This suggests that deviations from the midplane generally do occur but do not pervade the Galactic disk as they could be triggered by local perturbations.
   
   Wobbles and disk corrugations could be induced by merger events or satellite galaxies plunging through the Galactic plane as suggested by simulations of the stellar disk of galaxies \citep[e.g.,][]{1997MNRAS.287..947E,2014MNRAS.440.1971W,2017MNRAS.472.2751C}. This hypothesis is supported by stellar populations that have formed in the stellar disk but are found in the Galactic halo at $\sim$kpc distances from the plane \citep{2018Natur.555..334B}. Unsupervised machine learning measurements of stellar populations in the solar neighborhood reveal filamentary or string-like clusters that are oriented parallel to the Galactic disk, partly tens of parsecs above and below the plane \citep{2019AJ....158..122K}. Their filamentary shape is argued to be primordial and not an effect of tidal interaction, suggesting that the structure might be inherited from parental molecular clouds.
   
%-----------------------------------------------------------------

\section{Conclusions}\label{sec:conclusions}
   We have studied the atomic gas properties within the giant atomic filament Maggie. Kinematic information was obtained from the spectral decomposition of \ion{H}{i} emission spectra using the automated Gaussian decomposition routine \textsc{GaussPy+}. The main results are summarized as follows:
   
   \begin{enumerate}
      \item Maggie is one of the largest coherent filaments identified in the Milky Way, detected solely in atomic gas. At a kinematic distance of $\sim$17\,kpc, the projected length of $1.2\pm 0.1\rm\,kpc$ exceeds those of the largest molecular filaments identified to date. Maggie has an aspect ratio of 30:1, thus making its filamentary shape comparable to smaller-scale molecular filaments. Despite being parallel to the Galactic disk, Maggie has a distance of $\approx$500$\rm\,pc$ from the midplane. 
      \item The centroid velocity information obtained from the spectral decomposition reveals an undulating velocity structure with a smooth gradient of less than $3\rm\,km\,s^{-1}\,(10\,pc)^{-1}$. Based on the kinematic information, Maggie could be remotely linked to the global spiral structure of the Milky Way as it is trailing the Outer Arm by $5-10\rm\,km\,s^{-1}$. There is a slight bimodality in the line width distribution that is attributed to the phase degeneracy of the \ion{H}{i} emission and broad components blending in at similar $\varv_{\mathrm{LSR}}$ velocities.
      \item Optical depth measurements against strong continuum sources indicate that Maggie is composed of a significant fraction of CNM. Detections and non-detections of absorption toward extragalactic sources and Galactic \ion{H}{ii} regions, respectively, suggest that the distance inference based on the kinematic information is reasonable and that Maggie is located on the far side of the Galaxy. $K_\mathrm{S}$-band extinction obtained from 2MASS and WISE combined with Gaia parallaxes confirm that Maggie is not located within a distance of 5\,kpc from us. After correcting for optical depth effects, the column density has a mean of $\langle N_{\ion{H}{i}}\rangle=4.8\times 10^{20}\rm\,cm^{-2}$. We estimate a ``dense'' filament mass of $M=7.2\substack{+2.5 \\ -1.9}\times 10^5\,M_{\odot}$, providing a large atomic gas reservoir for molecular gas to form. If we assume radial symmetry about the major axis, the column densities equate to an average hydrogen number density of $\sim$4$\rm\,cm^{-3}$.
      \item We find no molecular counterpart to Maggie as traced by integrated \element[][]{CO} emission or dust. However, by summing diffuse CO emission voxels within \ion{H}{i} cloudlets obtained with a dendrogram analysis, \citet{2020PASJ...72...43N} confirm the presence of molecular gas on the smallest spatial scales. But as the molecular gas mass is only a few percent of the total mass budget of the filament, Maggie still appears to be in a predominantly atomic phase. Yet we note that \element[][]{CO} might not be a good probe at low column densities and early evolutionary stages as the onset of molecular filaments becoming CO-bright could lag behind the formation of $\rm H_2$. Although the formation timescale of molecular hydrogen in CNM-rich structures can be on the order of a few tens of Myr and Maggie exposes the onset of molecule formation in the highest-density pockets of \ion{H}{i}, the formation of $\rm H_2$ is not yet likely to be efficient in the more diffuse parts of the filament.
      \item Kinematic signatures are investigated by means of the Mach number distribution and reveal transonic and moderately supersonic velocities. The velocity structure function exhibits a modest power-law scaling that is still consistent with the Kolmogorov scaling for subsonic turbulence and is at the lower end of commonly observed scaling exponents in molecular clouds. The column density PDF can be represented by a log-normal distribution with a moderate width of $\sigma=0.28$, indicative of an intermediate CNM-WNM mix that is not dominated by the effects of gravitational contraction.
      \item Given the large distance from the Galactic midplane while remaining well aligned with the disk, we speculate that Maggie could be the signature of Galactic disk oscillations in the vertical direction. Future simulations of atomic and molecular structure formation in spiral galaxies under the influence of vertical perturbations could help constrain this possible formation pathway.
   \end{enumerate}
   Clearly, further work is needed to better understand the origins of Maggie. Higher-sensitivity CO observations as well as CO-dark molecular gas tracers will facilitate our understanding of this unique filament.
   While Maggie so far constitutes a case study, future studies of the entire THOR survey as well as other \ion{H}{i} surveys will investigate the more general nature of such large atomic filamentary structures.

\begin{acknowledgements}
      JS, HB, SCOG, and RSK acknowledge support from the Deutsche Forschungsgemeinschaft (DFG, German Research Foundation) -- Project-ID 138713538 -- SFB 881 (``The Milky Way System'', subprojects A01, B01, B02, and B08). HB and JDS further acknowledge funding from the European Research Council under the Horizon 2020 Framework Program via the ERC Consolidator Grant CSF-648505. This project has received funding from the European Research Council (ERC) under the European Union’s Horizon 2020 research and innovation programme (Grant agreement No. 851435). SB acknowledges the Institute for Theory and Computations (ITC) at the Harvard-Smithsonian Center for Astrophysics for financial support. This research was carried out in part at the Jet Propulsion Laboratory, California Institute of Technology, under contact with the National Aeronautics and Space Administration (80NM0018D0004). SCOG and RSK also acknowledge financial support from the Heidelberg Cluster of Excellence \mbox{STRUCTURES} in the framework of Germany’s Excellence Strategy (grant EXC-2181/1 - 390900948) and funding from the European Research Council via the ERC Synergy Grant ECOGAL (grant 855130). RJS is funded by an STFC ERF (grant ST/N00485X/1). This work has made use of data from the European Space Agency (ESA) mission Gaia (\url{https://www.cosmos.esa.int/gaia}), processed by the Gaia Data Processing and Analysis Consortium (DPAC, \url{https://www.cosmos.esa.int/web/gaia/dpac/consortium}). Funding for the DPAC has been provided by national institutions, in particular the institutions participating in the Gaia Multilateral Agreement. This research has made use of Astropy and affiliated packages, a community-developed core Python package for Astronomy \citep{2018AJ....156..123A}, Python package SciPy\footnote{\url{https://www.scipy.org}}, and APLpy, an open-source plotting package for Python \citep{aplpy2012}.\newline
      
      We thank the anonymous referee for the detailed comments and appreciate the effort that has clearly gone into reviewing our work. JS also thanks Mattia Sormani for the fruitful discussions that helped improve this paper.

\end{acknowledgements}
\bibliographystyle{aa_url} % style aa.bst

\bibliography{references} % your references Yourfile.bib

\onecolumn
\begin{appendix}
\section{Column density of Maggie component}\label{sec:coldens_app}
   \begin{figure*}
     \centering
     \includegraphics[width=1.0\textwidth]{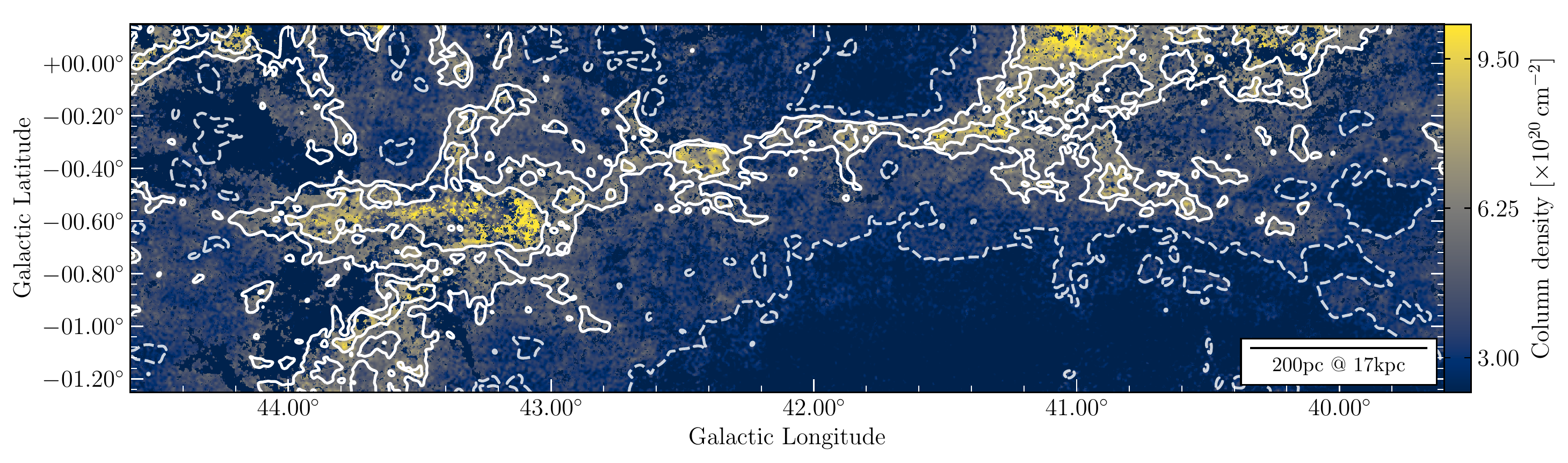}
        \caption[]{Integrated column density of the fitted Maggie component. In a thermally bistable medium, two components (i.e. CNM and WNM), which are close in velocity, are encoded in the emission profile. We identify the Maggie component based on the mean position in the velocity domain, thus picking up both CNM and WNM features. The contours are the same as in Fig.~\ref{fig:Maggie_massdens_map}.}
        \label{fig:coldens_Maggiecomp}
   \end{figure*}

\section{MWISP CO observations}\label{sec:MWISP_app}
   \begin{figure*}
     \centering
     \includegraphics[width=1.0\textwidth]{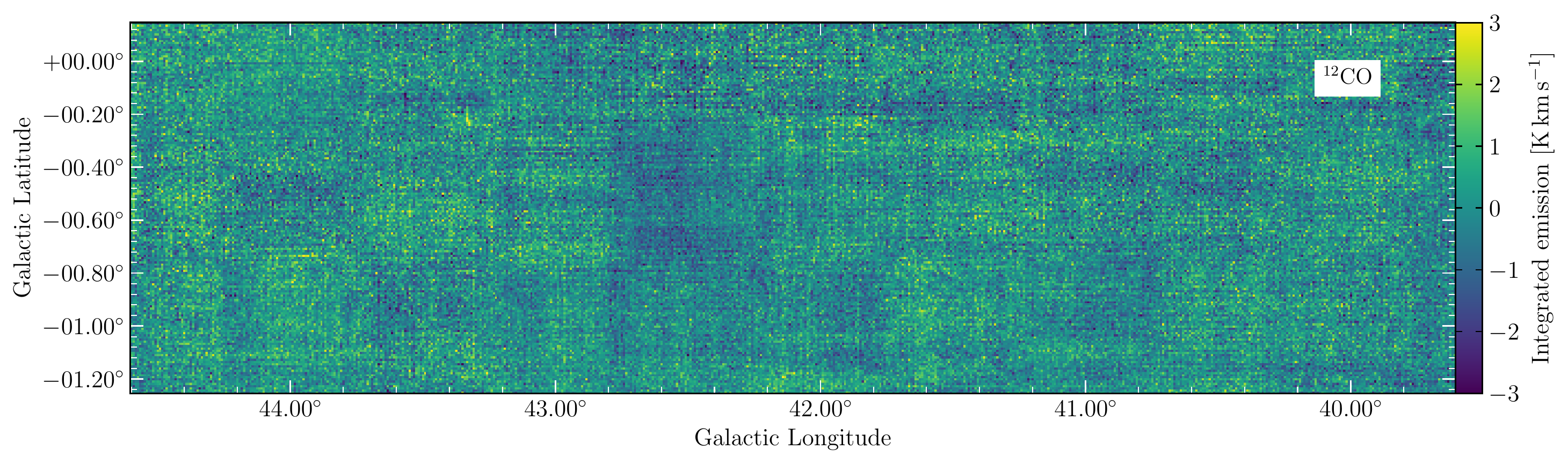}
        \caption[]{MWISP \element[][12]{CO} integrated emission. The \element[][12]{CO}\,($J=$1--0) emission data taken from the MWISP survey are integrated over the velocity interval between $-$57.5 and $-$48.5$\rm\,km\,s^{-1}$.}
        \label{fig:mwisp_map}
   \end{figure*}

\section{Stellar extinctions using Gaia, 2MASS, and WISE}\label{sec:Gaia_app}
   \begin{figure*}
     \centering
     \includegraphics[width=1.0\textwidth]{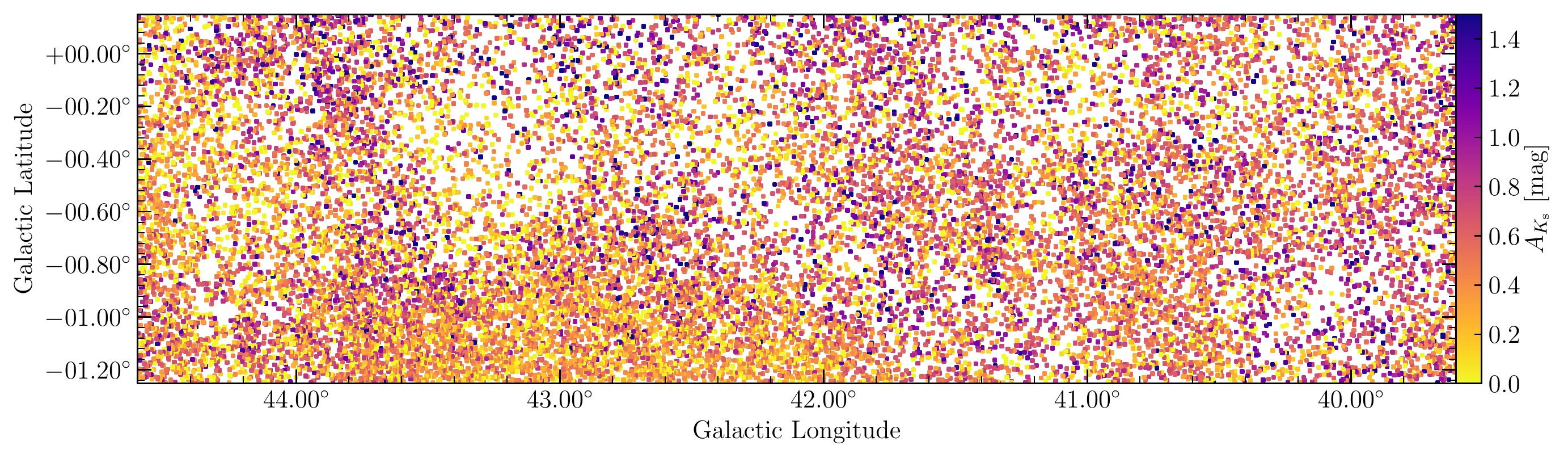}
        \caption[]{Extinctions up to 5\,kpc distance using \textit{Gaia} in combination with near- and mid-infrared photometry data from 2MASS and WISE. The blank spaces are regions containing obscuring material that has blocked out stars from our \textit{Gaia} sample and left some highly extinguished stars in their surroundings.}
        \label{fig:gaia_map}
   \end{figure*}
   
\end{appendix}

\end{document}